\documentclass[a4paper,12pt,twoside]{article}
\usepackage{a4,amsmath,amssymb,latexsym,booktabs}
\newcommand{\bm}[1]{\mbox{\boldmath{$#1$}}}
\newcommand{\bdgr}[1]{\bm{ #1}}
\newcommand{\sbdgr}[1]{\text{\bm{#1}}}
\newcommand{\secindent}{\hspace*{5mm}}
\newlength{\captionwidth}
\begin{document}

\thispagestyle{empty}

\begin{center}
{\large Optimal Design for Probit Choice Models with Dependent Utilities}\\[3ex]
%\end{center}
%\renewcommand{\thefootnote}{\alph{footnote}}
%\begin{center}
Ulrike Gra{\ss}hoff\,\footnote{Humboldt University Berlin, School of Business and Economics, Unter den Linden 6, D--10\,099 Berlin, Germany},
Heiko Gro{\ss}mann\footnote{Otto--von--Guericke University Magdeburg, Institute of Mathematical
Stochastics, PF 4120,
D--39\,016 Magdeburg, Germany}\\
Heinz Holling\;\!\footnote{University of M\"unster, Institute for Psychology,
Fliednerstr.\ 21,
D-48149 M{\"u}nster, Germany}, Rainer Schwabe\footnotemark[2]$^{,*}$
\renewcommand{\thefootnote}{\fnsymbol{footnote}}%
\footnotetext[1]{corresponding author,
{\tt e-mail: rainer.schwabe@ovgu.de}}
\\[3ex]
\end{center}
%\vfill
\begin{center}
%{\sc Abstract}
%\\[2mm]
\begin{minipage}{12cm}
{\begin{footnotesize} \secindent \end{footnotesize}}
\end{minipage}
\end{center}
%\vfill
\begin{abstract}
	In this paper we derive locally D-optimal designs for discrete choice experiments based on multinomial probit models. These models include several discrete explanatory variables as well as a quantitative one. The commonly used multinomial logit model assumes independent utilities for different choice options. Thus, D-optimal optimal designs for such multinomial logit models may comprise choice sets, e.g., consisting of alternatives which are identical in all discrete attributes but different in the quantitative variable. Obviously such designs are not appropriate for many empirical choice experiments. It will be shown that locally D-optimal designs for multinomial probit models supposing independent utilities consist of counterintuitive choice sets as well. However, locally D-optimal designs for multinomial probit models allowing for dependent utilities turn out to be reasonable for analyzing decisions using discrete choice studies.

\end{abstract}
{\begin{footnotesize}
\begin{description}
\item {\sc Keywords:} optimal design, paired comparison, discrete choice model, multinomial probit model
%\\[2ex]
\item
\\
\end{description}
\end{footnotesize}}
%\newpage

\newpage
\section{Introduction}
%\\[5mm]
%{\bf \large 1\ \ Introduction}
%\\[2mm]
Discrete choice analysis is a popular method for analyzing preferences and choices in economics, as well as in social and health sciences because it closely corresponds to making choices in everyday situations. In choice experiments respondents have to repeatedly choose between different alternatives within a so-called choice set. The alternatives also called options are defined by the levels of a subset of attributes. It is assumed that respondents choose the alternative with the greatest utility. The expected (overall) utility of an alternative is usually defined as a linear combination of part utilities assigned to the levels of the attributes of an alternative. The part and overall utilities are estimated from the choices of the respondents by using regression models.

Usually choice sets consist of two or three alternatives. When two alternatives are presented, discrete choice analysis coincides with paired comparison. Typically, the number of alternatives is held constant for all choice sets within a discrete choice experiment and all respondents will get the same series of choice sets, so the problem of designing the choice sets has to be considered for one respondent only.

Obviously, application of optimal design principles will be important to efficiently estimate the utilities represented by the parameters of the regression models. Usually, multinomial logit models have been applied to estimate the utilities. Several authors, see e.\,g.\ Gra{\ss}hoff et al., 2013; Kanninen, 2002), have developed optimal designs for discrete choice models based on multinomial logit models. However, the derived designs do not seem to be suitable for many empirical studies. E.g., when the set of attributes comprises several discrete attributes as well as a  further quantitative one, locally D-optimal optimal designs for such multinomial logit models consist of choice sets with alternatives that are identical in all discrete attributes but differ in the quantitative variable. This counterintuitive result is closely related to the assumption of “Independence from Irrelevant Alternatives” (IIA), characterizing logit regression. According to the IIA property, the choice probabilities of any two alternatives in a choice set are independent of all other alternatives contained in this choice set. However, such an assumption is inadequate for many everyday choice situations as the so-called Red-Bus/Blue-Bus Problem illustrates. Here, a subject can choose between two alternatives to get to work, say a bicycle and a red bus, each having a choice probability of .50. Consider now, in addition, a blue bus is as a third available option with identical attribute levels, except the attribute color and the part utilities for red and blue do not differ. According to the IIA property, the choice probabilities for the bicycle, red bus and blue bus then turn out to be .333 for each alternative. However, a multinomial probit model would lead to more reasonable choice probabilites of .50, .25 and .25 for the bicycle, the red and blues bus, respectively.

In the multinomial logit model, the utilities of the options follow a Gumbel-distribution. Furthermore, all utilities are mutually independent because of the IIA property. In this article we will analyze whether a multinomial probit model will also yield counterintuitive D-optimal designs when the utilities are independent.  Furthermore, we will derive such designs for multinomial probit models allowing for dependencies between the utilities of the alternatives. These models are based on assumptions which seem to be more realistic for most everyday choice situations.

The paper is organized as follows: In the next section two multinomial probit models will be introduced, one without dependent utilities and the other with dependency between the utilities. In section 3, we will derive locally D-optimal designs for both models including two alternatives, i.e., paired comparisons. First, the case of models including several qualitative attributes will be considered and then the more general case including a further quantitative attribute. In section 4, the results derived for paired comparisons will be generalized for both probit models including three options. The last section contains a short discussion of the results. All technical details are deferred to the Appendix.

\secindent
%%%%%%%%%%%%%%%%%%%%%%%%%%%%%%%%%%%%%%%
\section{Model description}
In a choice experiment individual choices are performed among $m\geq 2$ alternatives $\mathbf{a}_j$ of a choice set $\mathbf{A}=(\mathbf{a}_1,...,\mathbf{a}_m)$.
Each alternative $\mathbf{a}_j=(a_{j1},...,a_{jK})$, $j=1,...,m$ is characterized by $K$ attributes, where $a_{jk}$ is the level of the $k$th attribute presented in alternative $j$.
The decision behavior of a respondent can be described by a multinomial response $\mathbf{Y}=(Y_1,...,Y_m)^{\top}$, where $Y_j=Y_j(\mathbf{A})=1$, if $\mathbf{a}_j$ is chosen from $(\mathbf{a}_1,...,\mathbf{a}_m)$ and $Y_j=0$ otherwise, and $\mathbf{p}=\mathbf{p}(\mathbf{A})$ is the corresponding vector $\mathbf{p}=(p_1,...,p_m)^\top$ of probabilities of preference $p_j=p_j(\mathbf{A})=P(Y_j(\mathbf{A})=1)$ for the choice of the $j$th alternative $\mathbf{a}_j$ from a choice set $\mathbf{A}$.

These probabilities of preference are assumed to depend on latent utilities $U_j=U_j(\mathbf{a}_j)$ for all alternatives $\mathbf{a}_1,...,\mathbf{a}_m$ within the choice set $\mathbf{A}$, and the response is assumed to be obtained by the concept of utility maximization, i.\,e.\ $Y_j(\mathbf{A})=1$, if $U_j(\mathbf{a}_j) = \max_i U_i(\mathbf{a}_i)$.
Note that in general $P(U_i=U_j)=0$, as the utilities typically have continuous distributions, and, hence, the $Y_j$ are almost surely well defined.

In contrast to the commonly used multinomial logit choice model (see e.\,g.\ Gra{\ss}hoff et al., 2013, and the literature cited therein) we adopt here specifications of the latent utilities based on the normal distribution, which leads to a multinomial probit model.
This approach has the notable advantage that the utilities
\[
U_j(\mathbf{a}_j)=\sum_{k=1}^K U_{jk}(a_{jk})
\]
can be decomposed into part-worths $U_{jk}$ for the single attributes.
Within each alternative $\mathbf{a}_j$ the part-worth utilities $U_{jk}=U_{jk}(a_{jk})$ will be assumed to be independent, normally distributed with mean part-worths $\mu_{jk}=\mu_{jk}(\mathbf{a}_j)$, which depend only on the $k$th attribute each.
These mean part-worths $\mu_{jk}(\mathbf{a}_j)=\mathbf{f}_k(a_{jk})^{\top}\bdgr{\beta}_k$ are specified by linear effects with known regression function vectors $\mathbf{f}_k$ and unknown parameter vectors  $\bdgr{\beta}_k$ for each attribute $k$ separately.
Then the latent utility $U_j (\mathbf{a}_j)$ of an alternative $\mathbf{a}_j$ has mean $\mu_{j}=\mu_{j}(\mathbf{a}_j)=\mathbf{f} (\mathbf{a}_j)^{\top} \bdgr{\beta}$ with joint regression function $\mathbf{f}(\mathbf{a}_j) =(\mathbf{f}_1(a_{j1})^{\top},...,\mathbf{f}_K(a_{jK})^{\top})^{\top}$ and parameter vector
$\bdgr{\beta} = ( \bdgr{\beta}^{\top}_1 , ... , \bdgr{\beta}_K^{\top})^{\top}$, where $\mu_{j}=\sum_{k=1}^{K}\mu_{jk}$.
Typically the part-worth regression functions $\mathbf{f}_k$ will consist of dummy variables for qualitative factors, or they will be linear, if $a_{jk}$ is quantitative.

For simplification we will assume that all part-worth utilities share a common variance $\sigma_{0}^2$, i.\,e.\ $U_{jk}\sim N(\mathbf{f}_k(a_{jk})^{\top}\bdgr{\beta}_k,\sigma_{0}^2)$, throughout this paper, if not stated otherwise.

In what follows it will be crucial to specify the dependence structure between the $m$ utilities $U_{1},...,U_{m}$.
For this we  consider two particular models implied by different assumptions on the dependence between the part-worth utilities for an attribute $k$ across the alternatives.

\begin{tabular}{ll}
Model I: & all $U_{jk}$ and $U_{i \ell}$ are independent.
\end{tabular}

This model assumes independence of the part-worth utilities irrespectively whether the attributes of two alternatives differ or not and, thus,
results in the standard probit model considered in the literature, which may lead to counter-intuitive results similar to those for the common logit model (cf.\ Gra{\ss}hoff et al., 2013) as will be seen later.
To avoid these problems a second model is introduced, which accounts for dependence when the same level is presented for an attribute in different alternatives to be compared.

\begin{tabular}{ll}
Model II: & $U_{jk}=U_{ik}$, if $a_{jk}=a_{ik}$,\\
          & $U_{jk}$ and $U_{i  \ell}$ are independent, if $k\neq \ell$ or $a_{jk} \neq a_{ik}$.
\end{tabular}

In this model it is assumed that the presentation of equal levels for an attribute results in identical part-worth utilities in the alternatives presented together. Hence, in Model~II attributes with equal levels ($a_{ik}=a_{jk}$) will not contribute to the decision between alternatives $\mathbf{a}_i$ and $\mathbf{a}_j$ and the utilities $U_i$ and $U_j$ of the alternatives will become dependent.
% A corresponding decomposition into part utilities does not seem to be possible for the logit model.

Under the assumptions of Model~I as well as of Model~II the $m$-dimensional vector $\mathbf{U}=\mathbf{U}(\mathbf{A})=(U_1(\mathbf{a}_1),...,U_m(\mathbf{a}_m))^\top$ of utilities is multivariate normal with mean $\bdgr{\mu}(\mathbf{A})=(\mu_1(\mathbf{a}_1),...,\mu_m(\mathbf{a}_m)^\top$ and covariance matrix $\mathbf{V}(\mathbf{A})$.
In both models the utilities have equal variances $\mathrm{Var}(U_j)=\sigma_K^2=K\sigma_0^2$.
While in Model~I the utilities are independent such that $\mathbf{V}(\mathbf{A})=\sigma_K^2\mathbf{I}_m$, where $\mathbf{I}_m$ denotes the $m\times m$ identity matrix, the utilities become correlated in Model~II, when identical levels occur for some attributes.

According to the concept of utility maximization the alternative $j$ will be preferred to the other alternatives, if the utility $U_j$ is greater than all other utilities $U_i$, $i\neq j$. This implies for the preference probability
$$
\begin{array}{lll}
p_j = p_j (\mathbf{A}) = P(Y_j(\mathbf{A})=1) & = & P(\, U_j(\mathbf{a}_j)\geq \max_{i\neq j} U_i(\mathbf{a}_i) \, )\\
                                              & = & P(\, U_i (\mathbf{a}_i)- U_j(\mathbf{a}_j) \leq 0 \quad \mbox{for all} \quad i \neq j)
\end{array}
$$
For fixed $j$ let $\mathbf{L}_j$ the $(m-1)\times m$ matrix which transforms the $m$-dimensional vector $\mathbf{U}$ of utilities to the $(m-1)$-dimensional vector $\mathbf{U}_{(j)}=(U_i-U_j)_{i=1,...,m, i\neq j}$ of relevant utility differences ($\mathbf{U}_{(j)}=\mathbf{L}_j\mathbf{U}$).
Then the $(m-1)$-dimensional vector $\mathbf{U}_{(j)}(\mathbf{A})$ of utility differences $U_{i}(\mathbf{a}_{i})-U_j(\mathbf{a}_{j})$ is multivariate normal with mean vector $\bdgr{\mu}_j(\mathbf{A})=\mathbf{L}_j\bdgr{\mu}(\mathbf{A})$ and covariance matrix $\mathbf{V}_j(\mathbf{A})=\mathbf{L}_j\mathbf{V}(\mathbf{A})\mathbf{L}_j^\top$.

In any case the preference probability
\[
p_j(\mathbf{A})=\eta(\bdgr{\mu}_{j}(\mathbf{A}),\mathbf{V}_j(\mathbf{A}))
\]
can be written as a function of the mean vector $\bdgr{\mu}_j$ and the covariance matrix $\mathbf{V}_j$, where $\eta(\bdgr{\mu}_j,\mathbf{V}_j) = \Phi_{m-1} (\mathbf{0};\bdgr{\mu}_j,\mathbf {V}_j)$ denotes the distribution function of the $(m-1)$-dimensional normal variate with mean vector $\bdgr{\mu}_j$ and covariance matrix $\mathbf{V}_j$ evaluated at $\mathbf{0}$.

With this notation we can express the $m$-dimensional mean $E(\mathbf{Y}(\mathbf{A}))=\mathbf{p}(\mathbf{A})$ of the response $\mathbf{Y}(\mathbf{A})$ as
\[
E(\mathbf{Y}(\mathbf{A}))=\bdgr{\eta}_{\mathbf{A}}(\mathbf{F}(\mathbf{A})^{\top}\bdgr{\beta}) \, ,
\]
where $\mathbf{F}(\mathbf{A})=(\mathbf{f}(\mathbf{a}_1),...,\mathbf{f}(\mathbf{a}_m))$ is the $p\times m$-dimensional multivariate regression function and
\[
\bdgr{\eta }_{\mathbf{A}}(\bdgr{\mu})=(\eta(\mu_{1},\mathbf{V}_1(\mathbf{A})),...,\eta(\mu_m,\mathbf{V}_m(\mathbf{A})))^{\top}.
\]

The covariance matrix $\textrm{Cov}(\mathbf{Y})$ of the response vector $\mathbf{Y}$ is given by $\bdgr{\Sigma} =  \bdgr{\Sigma} (\mathbf{A}; \bdgr{\beta})         =\textrm{diag}(\mathbf{p})-\mathbf{p}\mathbf{p}^{\top}$, where $\textrm{diag}(\mathbf{p})$ is the $m\times m$ diagonal matrix with diagonal entries $p_j$, $j=1,...,m$.

Hence, both the mean response vector and the covariance matrix of $\mathbf{Y}$ depend on the parameter $\bdgr{\beta}$ only through the vector of linear effects $\bdgr{\mu}=\mathbf{F}(\mathbf{A})^{\top}\bdgr{\beta}$ and in addition on the $m$ covariance matrices $\mathbf{V}_j= \mathbf{V}_j (\mathbf{A})$, which only involve the choice set $\mathbf{A}$ presented.
Thus the observations may be interpreted as outcomes from an extended multivariate generalized linear model.

In this situation
the information for a choice set $\mathbf{A}$ can be calculated as
\[
\mathbf{M}(\mathbf{A};\bdgr{\beta})=\left(\frac{\partial \bdgr{\eta}_{\mathbf{A}}}{\partial \bdgr{\beta}}\right)^{\top} \bdgr{\Sigma }
(\mathbf{A};\bdgr{\beta}) \frac{\partial \bdgr{\eta}_{\mathbf{A}}}{\partial \bdgr{\beta}},
\]
where $\frac{\partial \sbdgr{\eta}_{\mathbf{A}}}{\partial \sbdgr{\beta}}$ denotes the $m\times p$ functional matrix of partial derivates of the $m-1$ components of $\bdgr{\eta}_{\mathbf{A}}$ with respect to the $p$ components of $\bdgr{\beta}$.
Remind that $\bdgr{\Sigma}$ as well as $\frac{\partial \sbdgr{\eta}_{\mathbf{A}}}{\partial \sbdgr{\beta}}$ depend on $\bdgr{\beta}$ only through $\mathbf{F}(\mathbf{A})^{\top}\bdgr{\beta}$.

The chain rule for the differentiation of multidimensional functions leads to
\[
\frac{\partial \bdgr{\eta}_{\mathbf{A}}(\mathbf{F}(\mathbf{A})^{\top}\bdgr{\beta})}{\partial \bdgr{\beta}}= \mathbf{J}_{\sbdgr{\eta}_{\mathbf{A}}}(\mathbf{F}(\mathbf{A})^{\top}\bdgr{\beta}) \mathbf{F}(\mathbf{A})^{\top}
\]
where $\mathbf{J}_{\sbdgr{\eta}_{\mathbf{A}}}(\bdgr{\bdgr{\mu}})$ is the Jacobian of the function $\bdgr{\eta}_{\mathbf{A}}$ evaluated at $\bdgr{\mu}$.
Thus the information matrix can be written as
\begin{eqnarray*}
\mathbf{M}(\mathbf{A};\bdgr{\beta})& = & \mathbf{F}(\mathbf{A}) \mathbf{J}_{\sbdgr{\eta}_{\mathbf{A}}}(\mathbf{F}(\mathbf{A})^{\top}\bdgr{\beta})  \bdgr{\Sigma }(\mathbf{A};\bdgr{\beta})^{-1} \mathbf{J}_{\sbdgr{\eta}_{\mathbf{A}}}(\mathbf{F}(\mathbf{A})^{\top}\bdgr{\beta}) \mathbf{F}(\mathbf{A})^{\top}\\
 & = & \mathbf{F}(\mathbf{A})\,\bdgr{\Lambda }(\mathbf{A};\bdgr{\beta})\,  \mathbf{F}(\mathbf{A})^{\top}  ,
\end{eqnarray*}
where $\bdgr{\Lambda }=\mathbf{J}_{\sbdgr{\eta}_{\mathbf{A}}}^{\top} \bdgr{\Sigma }^{-1} \mathbf{J}_{\sbdgr{\eta}_{\mathbf{A}}}$ denotes the $m\times m$ intensity matrix.

To tackle the problem of finding an optimal design, i.\,e.\ the best possible selection of choice sets, we will make use of the approximate design theory introduced by Kiefer (see e.\,g.\ Kiefer, 1974):
An approximate design $\xi$ on the set $\mathcal{X}$ of all choice sets consist of, say, $n$ different choice sets $\mathbf{A}_{i} = ( \mathbf{a}_{i1} ,..., \mathbf{a}_{im})$
with weights $w_i \geq 0$ and $\sum^n_{i=1}  w_i = 1$, representing the relative frequencies of replications.
The normalized per observation information matrix is defined by
\[
\mathbf{M}  (\xi ; \bdgr{\beta})
= \sum_{i=1}^n w_i \mathbf{M} ( \mathbf{A}_{i} ; \bdgr{\beta})
= \sum_{i=1}^n w_i \mathbf{F}  ( \mathbf{A}_{i}) \bdgr{\Lambda} ( \mathbf{A}_{i} ;
\bdgr{\beta} )  \mathbf{F} (  \mathbf{A}_{i} )^{\top}  .
\]
Note that for an exact design the usual information matrix equals $N$ times the normalized one, where $N$ is the total number of observations (presentations of choice sets).

To measure the quality of a design we will make use of the most common criterion of $D$-optimality, i.\,e.\ we are looking for designs $\xi^*$ that are locally $D$-optimal at $\bdgr{\beta}$, which maximize the determinant of the information matrix $\mathbf{M} (\xi ; \bdgr{\beta})$ (see e.\,g.\ Silvey, 1980).

%%%%%%%%%%%%%%%%%%%%%%%%%%%%%%%%%%%%%%%%%%%%%%%%%%%%%%%%%%%%%%%%%%%%%%%%%%%%%%%%%%%%%%%%%%%%%%%%%%%%%%%%%%
\section{Paired comparisons}
First we will focus on the particular case of $m=2$ alternatives, which represents the probit paired comparison model: The choices are performed between two alternatives $\mathbf{a}_1$ and $\mathbf{a}_2$ of a pair $\mathbf{A}=(\mathbf{a}_1,\mathbf{a}_2)$.
Because of $Y_2=1-Y_1$ and $p_2=1-p_1$ we actually have to deal with only one preference probability $p = p_1$ for the first alternative in a pair.
The mean  of the binomial response variable $Y = Y_1$ is given by a one-dimensional function $\eta = \eta_1$, which leads to an extended generalized linear model with
\[
E(Y(\mathbf{A}))=\eta(\tilde{\mathbf{f}}(\mathbf{A})^{\top}\bdgr{\beta} , \sigma^2(\mathbf{A}))=\Phi_0( \tilde{\mathbf{f}}(\mathbf{A})^{\top}\bdgr{\beta}/\sigma ( \mathbf{A} )) \, ,
\]
where $\tilde{\mathbf{f}}(\mathbf{A})=\mathbf{f}(\mathbf{a}_2)-\mathbf{f}(\mathbf{a}_1)$, $\Phi_0 $ denotes the standard normal distribution function and the variance $\sigma^2(\mathbf{A})=\textrm{Var}(U_1(\mathbf{a}_1)-U_2(\mathbf{a}_2))$ is the one-dimensional counterpart of the covariance matrix $\mathbf{V}_1 (\mathbf{A})$. The variance of the response is given by $\mathrm{Var} (Y( \mathbf{A})) = p ( \mathbf{A}) \, (1 - p ( \mathbf{A}))$.
In the present case we have for the derivative
\[
\frac{\partial \eta(\tilde{\mathbf{f}}(\mathbf{A})^{\top}\bdgr{\beta} , \sigma^2(\mathbf{A}))}{\partial \bdgr{\beta}}=\frac{\varphi_0(\tilde{\mathbf{f}}(\mathbf{A})^{\top}\bdgr{\beta}/\sigma(\mathbf{A}))}{\sigma(\mathbf{A})}  \tilde{\mathbf{f}}(\mathbf{A})^{\top} ,
\]
where $\varphi_0$ is the density of the standard normal distribution.
Hence, the information for a pair $\mathbf{A}$ is given by
\[
\mathbf{M} (\mathbf{A};\bdgr{\beta})=\lambda(\mathbf{A};\bdgr{\beta})\, \tilde{\mathbf{f}}(\mathbf{A})\tilde{\mathbf{f}}(\mathbf{A})^{\top}
\]
with intensity function
\[
\lambda(\mathbf{A};\bdgr{\beta})= \frac{\varphi_0(\tilde{\mathbf{f}}(\mathbf{A})^{\top}\bdgr{\beta}/\sigma(\mathbf{A}))^2} {\sigma^2(\mathbf{A})\Phi_0(\tilde{\mathbf{f}}(\mathbf{A})^{\top}\bdgr{\beta}/\sigma(\mathbf{A}))(1-\Phi_0(\tilde{\mathbf{f}}(\mathbf{A})^{\top}\bdgr{\beta}/\sigma(\mathbf{A}))} \; ,
\]
which depends on $\bdgr{\beta}$ only through the linear component $\tilde{\mathbf{f}}(\mathbf{A})^{\top}\bdgr{\beta}$ and additionally on the scaling factor $\sigma(\mathbf{A})$.

%%%%%%%%%%%%%%%%%%%%%%%%%%%%%%%%%%%%%%%%%%%%%%%%%%%%%%%%%%%%%%%%%%%%%%%%%%%%%%%%%%%%%%%%%%%%%%%%%%%%%%%%%
\subsection{Qualitative attributes in the case of indifference}
To start we consider in this subsection the special case
$\bdgr{\beta}= \mathbf{0}$, which results in equal choice probabilities $p = 1 - p = 1/2$ for any pair of alternatives, which can be interpreted as the situation of indifference.

Under the assumption of Model~I we have constant variance $\sigma^2 = \sigma^2(\mathbf{A})=2K\sigma_0^2$ for all pairs $\mathbf{A}$. Then for an approximate design $\xi$ the information matrix
\[
\mathbf{M}(\xi; \mathbf{0}) = \frac{1}{\pi  K \sigma^2_0}\mathbf{M}_L(\xi)
\]
is proportional to the information matrix $\mathbf{M}_L(\xi) = \sum_{i=1}^n w_i \tilde{\mathbf{f}}(\mathbf{A}_i)\tilde{\mathbf{f}}(\mathbf{A}_i)^{\top}$ in the corresponding linear paired comparison model
(see Gra{\ss}hoff et al., 2004).
As a consequence any $D$-optimal design in the linear paired comparison model is also $D$-optimal in the probit paired comparison model, when
all utility terms $U_{1k}$ and $U_{2 \, \ell}$ are assumed to be independent and $\bdgr{\beta}=\mathbf{0}$.

Under the assumptions of Model~II the comparison depth $d_{\mathbf{A}}$ will play an important role, where $d_{\mathbf{A}} = \# \, \{k; a_{1k} \neq a_{2k}\}$ is defined as the number of attributes, for which the components differ within the pair
$\mathbf{A} = (\mathbf{a}_1 ; \mathbf{a}_2)$.
With this notation the variance of the utility difference can be written as $\sigma^2(\mathbf{A})=2d_{\mathbf{A}}\sigma_0^2$.

To simplify the problem further we consider a setting of $K$ qualitative factors, which may be adjusted to the same number $v_k=v$ of levels $ 1,...,v$, say, for each attribute $k$. The vector $\mathbf{f}=(\mathbf{f}_1^\top,...,\mathbf{f}_K^\top)^\top$ of
part-worth regression functions is chosen according to effect coding. More precisely, $\mathbf{f}_k(i)=\mathbf{e}_{v-1;i}$, if $i=1,...,v-1$, where $\mathbf{e}_{v-1;i}$ denotes the $i$th unit vector of length $v-1$,
and $\mathbf{f}_k (v) = - \mathbf{1}_{v-1}$, where $\mathbf{1}_{v-1}$ denotes the vector of length $v-1$ with all entries equal to $1$ (for more details on this model specifications see Gra{\ss}hoff et al., 2004).

Since in the present situation the $D$-criterion is invariant with respect to both permutations of the levels for each attribute and to permutations of the attributes themselves,
optimal designs can be found within the class of invariant designs which are uniform on the orbits induced by these permutations. These orbits are the sets of pairs with a fixed comparison depth $d \leq K$.

By $\bar{\xi}_{d}$ we denote the design which is uniform on the orbit of comparison depth $d$.  In particular, for full comparison depth
$d=K$ the uniform design $\bar{\xi}_K$ is the product type design $\xi_0 \otimes \ldots \otimes \, \xi_0$, where $\xi_0$ is the uniform balanced incomplete block design with blocks of size $2$ consisting of the $v(v-1)$ pairs concerning one single attribute.
Thus $\bar{\xi}_K$ is uniform on all pairs, which have different levels in each attribute. Gra{\ss}hoff et al. (2004) established that the design $\bar{\xi}_K$ is $D$-optimal
in the linear paired comparison model and, thus, it is also optimal in Model~I.
In that case the optimal information matrix equals
$\mathbf{M} (\bar{\xi}_K ; \mathbf{0}) = ( \mathbf{I}_K\otimes \mathbf{M}^* )/(\pi K \sigma_0^2)$,
where $\mathbf{M}^* = \frac{2}{v-1}(\mathbf{I}_{v-1}+\mathbf{1}_{v-1}\mathbf{1}_{v-1}^{\top})$ is the information matrix of the marginal design $\xi_0$ in the single attribute linear paired comparison model and ``$\otimes$'' is the symbol for the Kronecker product of matrices.

In contrast to that under the assumptions of Model~II for pairs belonging to the orbit of comparison depth $d \geq 1$ the intensity $1/(\pi d \sigma_0^2)$ depends on the comparison depth $d$. Simple combinatorial arguments lead to the information matrix
\[
\mathbf{M} (\bar{\xi}_d ; \mathbf{0}) = \frac{d}{K}\,\frac{1}{\pi  d \sigma^2_0}( \mathbf{I}_K\otimes \mathbf{M}^*)
=\mathbf{M}(\bar{\xi}_K ; \mathbf{0}) \, .
\]
Note that for $d=0$ all attributes and, hence, both alternatives and their corresponding utilities completely coincide.
Therefore the resulting information is equal to zero ($\mathbf{M}(\bar{\xi}_0;\mathbf{0})=\mathbf{0}$).

Since $\mathbf{M}  (\bar{\xi}_d ; \mathbf{0})$ is independent of the comparison depth $d\geq 1$, all designs  $\bar{\xi}_d$ and, in particular,
the design $\bar{\xi}_K$, which is $D$-optimal under Model~I, are also $D$-optimal under Model~II.
Furthermore, any convex combination of the designs $\bar{\xi}_d $, $d\geq 1$, is also $D$-optimal under Model~II.

%%%%%%%%%%%%%%%%%%%%%%%%%%%%%%%%%%%%%%%%%%%%%%%%%%%%%%%%%%%%%%%%%%%%%%%%%%%%%%%%%%%%%%%%%%%%%%%%%%%%%%%%55
\subsection{One additional quantitative attribute}
We extend the model to the situation investigated by Kanninen (2002), which led to counter-intuitive results in the logit model after design optimization for larger choice sets (see Gra{\ss}hoff et al., 2013) and which caused us to introduce Model~II.

The purpose of the present subsection is to provide optimal designs for probit paired comparison models with and without dependence structure in the part-worth utilities before studying larger choice sets.
More precisely, we consider a model with pairs
$\mathbf{A}=(\mathbf{a}_1,\mathbf{a}_2)$ of alternatives, where one of the attributes, say the last one, is quantitative and unrestricted and can be interpreted, for example, as a price variable (potentially on a logarithmic scale) and all other attributes are qualitative.
Then the set of attributes can be split into two components
$\mathbf{a}_j=(\mathbf{x}_j^{\top},t_j)^{\top}$, where $t_j \in {\Bbb R}$ and $\mathbf{x}_j$ consists of the qualitative attributes. According to the marginal pairs $\mathbf{x}=(\mathbf{x}_1,\mathbf{x}_2)$ and $\mathbf{t}=(t_1,t_2)$ we can decompose the regression function for $Y_1$ as
\[
\tilde{\mathbf{f}}(\mathbf{A}) = (\tilde{\mathbf{f}}_1(\mathbf{x})^{\top},\tilde{f}_2(\mathbf{t}))^{\top} \, ,
\]
where the marginal regression functions are defined by $\tilde{\mathbf{f}}_1(\mathbf{x})=\mathbf{f}(\mathbf{x}_1)-\mathbf{f}(\mathbf{x}_2)$, $\tilde{f}_2(\mathbf{t})=t_1-t_2$, and for the qualitative attributes the regression function $\mathbf{f}$ is defined as in subsection 3.1.

Following Kanninen (2002) we restrict our investigations for the first component to the setting of $K$ binary attributes, varying on $v = 2$ levels each, i.\,e.\ $\mathbf{x}_j=(x_{j1},...,x_{jK})^{\top}\in \{1,2\}^K$.
Under effect coding the corresponding regression
functions are given by $\mathbf{f}(\mathbf{x}_j)=(f_k(x_{jk}))_{k=1,...,K}$
with $f_k(1) = 1$ and $f_k(2) = -1$.

The utility $U_j(\mathbf{a}_j )$ in this two component model is generated by partial utilities
\[
U_j(\mathbf{a}_j) = U_{j1}(\mathbf{x}_j)+U_{j2}(t_j) \, ,
\]
where the partial utility  $U_{j1}(\mathbf{x}_j)=\sum_{k=1}^K U_{j1k}(x_{jk})$ of the first component is itself composed of part-worth utilities $U_{j1k}$, which are assumed to be independent and normally distributed
with mean $f_{k}(x_{jk})\beta_{1k}$ and constant variance $\sigma_0^2$ across the attributes as in the previous subsection. For the second
component we assume a normally distributed part-worth utility $U_{j2}$ with mean $\beta_2 \, t_j$
and variance $\sigma^2_t \geq 0$, which is independent of the part-worth utilities $U_{j1k}$ of the first component. Furthermore we will assume throughout that all part-worth utilities $U_{j2}$ for the second component are independent. As a special case we may allow for a sharp decision with respect to the quantitative attribute by letting $\sigma^2_t = 0$, which results in a degenerate utility $U_{j2} \equiv \beta_2 t_j$.

Optimal designs for such  a two component model were first investigated numerically by Kanninen (2002) in the binomial logit model.
Gra{\ss}hoff et al.~(2007) gave explicit proofs for $D$-optimal designs
by making use of a canonical transformation introduced by Ford et al.~(1992) and extended by Sitter and Torsney (1995) to the multifactorial case.
We will apply this construction method also to the probit models considered here.

To this end in a first step the standardized case $\bdgr{\beta_1=0}$ and $\beta_2=1$ is considered.
There the intensity function $\lambda$ for a pair $\mathbf{A}$ reduces to
$\lambda(\mathbf{A};\bdgr{\beta}) =  \lambda_2 ((t_{1}-t_{2})/\sigma (\mathbf{A}))/\sigma^2(\mathbf{A})$,
where $\lambda_2(z) = \varphi_0(z)^2/(\Phi_0 (z)(1-\Phi_0 (z)))$ is the marginal intensity with respect to the quantitative attribute.
Hence, the intensity $\lambda (\mathbf{A} ; \beta)$ depends on the first component $\mathbf{x}$ only through the scaling factor $\sigma (\mathbf{A})$.

The situation of independent utilities of Model~I results in the standard probit model in the literature:
If the part-worth utilities $U_{j1}$ of the first components satisfy the assumptions of Model~I, then $ \sigma(\mathbf{A}) = \sigma_{\max}$  attains the same value for all pairs $\mathbf{A}$, where $\sigma^2_{\max} = 2  (\sigma_K^2+\sigma_t^2)$ and, again, $\sigma_K^2=K\sigma_0^2$.
Thus, the intensity function only depends on the linear response through
the second component, and the approach described in Gra{\ss}hoff et al.\ (2007) can be used.

Denote by $\delta_{\,\mathbf{t}}$ the one-point design at $\mathbf{t} = (t_1 , t_2)$.
\\[3mm]
\textbf{Theorem 1. } \textit{Let $z^* > 0$ maximize $\lambda_2(z)^{K+1}z^2$ and let $\mathbf{t}^*$ satisfy $t^*_1 - t^*_2 = \sigma_{\max}  z^*$.
Then the design $\xi^*=\bar{\xi}_K \otimes \delta_{\,\mathbf{t}^*}$ is locally $D$-optimal at $\bdgr{\beta}=(\mathbf{0},1)^{\top}$ in the probit paired comparison model with independent part-worth utilities (Model~I).}
\vspace{3mm}

Table 1 lists the optimal values $z^*$ together with the corresponding preference probabilities $p=\Phi_0 (z^*)$ for various numbers $K$ of attributes for the first component.
\begin{table}
\begin{center}
\begin{tabular}{cccccccc}
$K$ & \phantom{??}  1 \phantom{??}  & \phantom{??}  2  \phantom{??} & \phantom{??}  4  \phantom{??} & \phantom{??}  8 \phantom{??}  &  \phantom{??} 10  \phantom{??} &  \phantom{??} 50  \phantom{??} &  \phantom{??} 100 \phantom{??} \\
\hline
$z^*$ & 1.138 & 0.938 & 0.732 & 0.549  & 0.497 & 0.232 & 0.165 \\
$\Phi (z^*)$ & 0.872 & 0.826 & 0.768 & 0.708 & 0.690 & 0.592 & 0.566
\end{tabular}
\caption{Optimal values $z^*$ and preference probabilities $ \Phi_0 (z^*)$}
\end{center}
\end{table}
Note that $z^*$ may be replaced by $-z^*$ and the optimal $z^*$ in the case $K = 1$ coincides with the optimal value for the standard probit regression model (see Ford et al., 1992).

%\begin{center}
%\begin{tabular}{cccc}
%$K$  & \vline & $z^*_K$ & $q(z^*_K)$  \\
%\hline
%1 & \vline & 1.138 & 0.872 \\
%2 & \vline & 0.938 & 0.826\\
%3 & \vline & 0.816 &0.793\\
%4 & \vline & 0.732 &0.768\\
%     5 & \vline & 0.670 &0.749\\
%    6 & \vline & 0.621 &0.733\\
%    7 & \vline & 0.581 &0.719\\
%    8 & \vline & 0.549 &0.708\\
%    9 & \vline & 0.521 &0.699\\
%   10 & \vline & 0.497 &0.690\\
%    20 & \vline & 0.361 &0.641\\
%    50 & \vline & 0.232 &0.592\\
%   100 & \vline & 0.165 &0.566\\
%\\
%\end{tabular}\\
%\footnotesize{Table 1: Optimal values $z^*_K$ and preference probabilities $q \, (z^*_K) = \Phi (z^*_K)$}
%\end{center}
%%%%%%%%%%%%%%%%%%%%%%%%%%%%%%%%%%%%%%%%%%%%%%%%%%%%%%%%%%%%%%%%5
As the model with independent utilities may lead to counter-intuitive results, if larger choice sets are considered, we introduce
a two component model, where the first component fulfills the assumptions of Model~II, but the part-worth utilities $U_{12}$ and $U_{22}$ will still be assumed to be independent (potentially degenerate). Then the scaling factor $\sigma(\mathbf{A})$
is obtained by
\[
\sigma^2(\mathbf{A})=2\, (d\, \sigma_0^2+\sigma_t^2) = (d\,\sigma^2_{\max} + (K-d) \, 2 \sigma_t^2)/K
\]
for pairs $\mathbf{A}$ belonging to an orbit of comparison depth $d$ in the qualitative attributes, where $\sigma^2_{\max} = 2 (K\sigma_0^2+\sigma_t^2)$ is the maximal possible variance, which is achieved, if $\mathbf{A}$ has comparison depth $K$.
%If we rewrite the scaling factor as $2 \sigma_t^2=\varepsilon \sigma^2_0$, this implies $2 \sigma_0^2=\frac{1-\varepsilon}{K} \sigma^2_0$. Then the case $\varepsilon=0$ can be interpreted as  a ``sharp" decision concerning the price variable $\mathbf{t}$, as there is no variability in the corresponding part-worth utility. The limiting case $\varepsilon =1$ represents the situation of independent utilities treated before. With this notation we get
%\[
%\sigma(\mathbf{A})^2 = ((1 - \varepsilon) \, \frac{\D d}{\D K} + \varepsilon) \, \sigma^2_0 \, .
%\]
Irrespectively of the variation structure characterized by
%$\varepsilon$
$\sigma_0$ and $\sigma_t$
the optimal design of Theorem~1 also turns out to be optimal, here.
\\[3mm]
\textbf{Theorem~2. } \textit{If $z^* > 0$ maximizes $\lambda_2(z)^{K+1} z^2$ and if $\mathbf{t}^*$ satisfies $t^*_1 - t^*_2 = \sigma_{\max}  z^*$ then the design $\bar{\xi}_K \otimes \delta_{\,\mathbf{t}^*}$
is $D$-optimal for the probit paired comparison model with dependent utilities (Model~II).}
%%%%%%%%%%%%%%%%%%%%%%%%%%%%%%%%%%%%%%%%%%%%%%%%%%%%%%%%%%%%%%
\vspace{3mm}

In the general two component model with arbitrary $\bdgr{\beta}_1$ and $\beta_2$ we have to suppose $\beta_2\not= 0$
in order to guarantee the existence of a finite solution of the design optimization problem.
According to Gra{\ss}hoff et al.\ (2007) $D$-optimal designs can be constructed by using the concept of canonical transformations (see Ford et al., 1992, and Sitter and Torsney, 1995). The procedure is based
on a one-to-one mapping $g$ defined by $g(\mathbf{a}_j) = (\mathbf{x}^{\top}_j  , \, \mathbf{f}_1  (\mathbf{x}_j)^{\top}  \bdgr{\beta}_1 + t_j  \beta_2)^{\top}$
on the alternatives, which transforms to the case of indifference for the qualitative attributes.
The simultaneous transformation $\mathbf{g} (\mathbf{A}) = (g(\mathbf{a}_1)  , \, g (\mathbf{a}_2))$ of both alternatives induces
a linear transformation $\tilde{\mathbf{f}}(\mathbf{g} (\mathbf{A}))=\mathbf{Q}_g\tilde{\mathbf{f}}(\mathbf{A})$ of the induced regression functions with
\[
\mathbf{Q}_g=\left(
\begin{tabular}{cc}
$\mathbf{I}_K$ & $\mathbf{0}$\\
$\bdgr{\beta}_1^{\top}$ & $ \beta_2$
\end{tabular}
\right) \, .
\]

If we let  $z_j = \mathbf{f}_1(\mathbf{x}_j)^{\top} \bdgr{\beta}_1 + t_j  \beta_2$ for the unrestricted quantitative component in the transformed model, the information matrix coincides with the standardized
situation $\bdgr{\beta}_1 = \mathbf{0}$ and $\beta_2 = 1$. Then optimal designs can be obtained by a back transformation of the optimal design
$\bar{\xi}_K \otimes \delta_{\,\mathbf{t}^*}$ for the standardized situation: The induced design defined by $\xi^* (\mathbf{x} , \mathbf{t}) = \bar{\xi}_K (\mathbf{x})\,  \delta_{\,\mathbf{t}^*}(\mathbf{g} (\mathbf{x} , \mathbf{t}))$
turns out to be $D$-optimal, which establishes the following result.
\\[3mm]
\textbf{Theorem 3. } \textit{Let $z^*$ maximize $\lambda_2(z)^{K+1} z^2$. Denote  by $\xi^*_{2 | 1}$ the conditional design, which is concentrated on
$\mathbf{t}^*(\mathbf{x})$ for every pair $\mathbf{x}$, where $\mathbf{t}^*(\mathbf{x})=(t^*_1(\mathbf{x}),t^*_2(\mathbf{x}))$ satisfies $t^*_1(\mathbf{x})-t^*_2(\mathbf{x})= (\sigma_{\max}  z^*-\tilde{\mathbf{f}}_1(\mathbf{x})^{\top}\bdgr{\beta}_1)/\beta_2$. Then the combined
design $\xi^*=\bar{\xi}_K\otimes \xi_{2 | 1}^*$ is $D$-optimal under both model assumptions~I and II of independent or dependent utilities, respectively.}
\vspace{3mm}

If is worth-while mentioning that also in the general case the optimal values $\mathbf{t}^*(\mathbf{x})$ for the second component are chosen in such a way that the optimal preference probabilities
$p=P(Y (\mathbf{A}) = 1) = \Phi_0 (z^*)$ of Table~1 are retained.
%%%%%%%%%%%%%%%%%%%%%%%%%%%%%%%%%%%%%%%%%%%%%%%%%%%%%%%%%%%%%%%%%%%%%%%%%%%%%%

\section{Choice sets with three alternatives}
We turn now to the situation of choice sets with $m=3$ alternatives.
In contrast to paired comparisons there a reduction to one dimension is no longer possible, and we have to deal with proper multinomial observations.
To compute the preference probabilities $p_j$ for a choice set $\mathbf{A}=(\mathbf{a}_1,\mathbf{a}_2,\mathbf{a}_3)$ we use of the software package \texttt{mvtnorm} implemented in \texttt{R} (see Genz and Bretz, 2009, and Genz et al., 2017) for obtaining the multivariate normal probabilities in the variance terms.

For abbreviation we denote by $\sigma_{ij}^2(\mathbf{A})=\sigma_{ji}^2(\mathbf{A})$ the diagonal elements $\textrm{Var}(U_i-U_j)$ of the covariance matrix $\mathbf{V}_j$ and introduce the standardized mean differences
\[
z_{ij}(\mathbf{A})=((\mathbf{f}_{i}(\mathbf{a}_i)-\mathbf{f}_{j}(\mathbf{a}_j))^{\top}\bdgr{\beta})/\sigma_{ij}(\mathbf{A}) \,.
\]
Further let $\bdgr{\Phi}_{\varrho}$ be the bivariate normal distribution function with location vector zero, scaling parameters one and correlation coefficient $\varrho$ and denote by where $\varrho_{\,j}(\mathbf{A})=\textrm{corr}\,(U_i-U_j,U_\ell-U_j)$ the correlation in the covariance matrix $\mathbf{V}_j$.
With this notation  the preference probabilities can be rewritten as
\[
p_j(\mathbf{A})=\bdgr{\Phi}_{\varrho_j(\mathbf{A})} (z_{ji}(\mathbf{A}) ,\,z_{j\ell}(\mathbf{A})) \, ,
\]
where the indices $i$ and $\ell$ denote the other alternatives besides $j$.
Then the Jacobian matrix $\mathbf{J}_{\sbdgr{\eta}_{\mathbf{A}}}$ can be computed as
\[
\mathbf{J}_{\sbdgr{\eta}_{\mathbf{A}}}(\mathbf{F}(\mathbf{A})^{\top}\bdgr{\beta})=\left(
\begin{tabular}{ccc}
$h_{12}+h_{13} $ & $-h_{12}$ & $-h_{13}$
\\
$-h_{21}$ & $h_{21}+h_{23}$ & $-h_{23}$
\\
$-h_{31}$ & $-h_{32}$ & $h_{31}+h_{32}$
\end{tabular}
\right) \, ,
\]
where
\[
h_{ij} =h_{ij}(\mathbf{A}) = \varphi_0 (z_{ij}) \Phi_0((z_{i\ell}-\varrho_{i}z_{ij})/(1-\varrho\,_i^2)^{1/2})/\sigma_{ij}(\mathbf{A})
\]
and $\ell$ is the index of the third alternative besides $i$ and $j$.

% % % % % % % % % % % % % % % % % % % % % % % % % % % % % % % % % % % % % % % % % % % % % % % % % % % % % % %
\subsection{Qualitative attributes in the case of indifference}
Also here we first consider the particular case $\bdgr{\beta}=\mathbf{0}$ of indifference for the setting of $K$ qualitative attributes as in the corresponding subsection on paired comparisons.
However, for simplification we additionally restrict here to the case of $v = 2$ levels for each attribute.

As will be seen the intensity matrix $\bdgr{\Lambda}$ will not be affected under indifference and the assumptions of Model~I and II, respectively, when levels are permuted within attributes and attributes are permuted with each other.
Then also the $D$-criterion is invariant with respect to these permutations.
 Hence, as in the paired comparison case optimal designs can be found within the class of invariant designs, which are uniform on the orbits induced by the permutations.

 In order to characterize these orbits we introduce a multivariate analogue to the concept of comparison depth for paired comparisons.
 For any choice set $\mathbf{A}=(\mathbf{a}_1,\mathbf{a}_2,\mathbf{a}_3)$ we denote by $d_{ij}=d_{ij}(\mathbf{A})$ the number of attributes, for which the levels of the alternatives $\mathbf{a}_i$ and $\mathbf{a}_j$ differ, i.\,e.\ $d_{ij}$
is the comparison depth of the pair $(\mathbf{a}_i,\mathbf{a}_j)$
The triple $\mathbf{d}=\mathbf{d}(\mathbf{A})=(d_{12},d_{13},d_{23})$ will be called the comparison depth of the choice set.
Note that each attribute contributes either zero to the comparison depth the case that all alternatives coincide in this attribute, or it adds $1$ to two components of the comparison depth vector $\mathbf{d}$ in the situation that two alternatives are equal and the third one differs in this attribute.
Thus it is easy to see that the mean comparison depth $D=(d_{12}+d_{13}+d_{23})/2$ satisfies $D\leq K$.

In the following we will only consider choice sets with full profiles, for which the mean comparison depth is maximal ($D=K$), as choice sets with partial profiles ($D<K$), for which, at least, one attribute is equal across all alternatives, tend to bear less information (see Gra{\ss}hoff et al., 2009, for the logistic case).

All orbits are characterized by their comparison depth $\mathbf{d}$.
Because a permutation of the arrangement of the alternatives within a choice set does not affect the corresponding information matrix, an orbit described by the comparison depth $\mathbf{d}$ can be considered as being equivalent to an orbit associated with a permutation of the entries in $\mathbf{d}$:
For example in the situation of two identical alternatives the orbit $\mathbf{d}=(K,K,0)$ indicates that alternative $2$ equals alternative $3$, whereas on the orbit $\mathbf{d}=(K,0,K)$ alternative $1$ is equal to alternative $3$ and on $\mathbf{d}=(0,K,K)$ alternative $1$ and $2$ are coincide, while in each case the third alternative differs in all attributes.
Hence, without loss of generality we need only consider comparison depths satisfying $d_{12}\geq d_{13}\geq d_{23}$.

For the uniform design $\bar{\xi}_{\mathbf{d}}$ on the orbit $\mathbf{d}=(d_{12},d_{13},d_{23})$
the information matrix
\begin{equation}
\label{lamxid}
\mathbf{M}(\bar{\xi}_{\mathbf{d}})=4\, \lambda_{\mathbf{d}} \, \mathbf{I}_{K}
\end{equation}
is a multiple of the identity matrix.
The diagonal elements are given by the mean intensity
\[
\lambda_{\mathbf{d}}= \frac{1}{2K}\sum_{j=1}^{3}(d_{ji}+d_{j\ell}-d_{i\ell})\lambda_{jj}(\mathbf{d})
                                   = \frac{1}{K}\sum_{j=1}^{3}(K-d_{i\ell})\lambda_{jj}(\mathbf{d}) \, ,
\]
where, also here, the indices $i$ and $\ell$ denote the other alternatives besides $j$ and the $\lambda_{jj}(\mathbf{d})$ are the diagonal entries of the intensity matrix $\bdgr{\Lambda}(\mathbf{d})$ on the orbit $\mathbf{d}$.
Note that for the off-diagonal entries of $\mathbf{\Lambda}$ the relation $2\lambda_{ij}=\lambda_{\ell\ell}-\lambda_{ii}\lambda_{jj}$ holds.
The determinant $\det \mathbf{M} (\xi_{\mathbf{d}})$ of the information matrix will then be maximized by the uniform design on the orbit $\mathbf{d}$, which yields the largest value of $\lambda_{\mathbf{d}}$.

Under the assumption of Model~I we observe that the variances $\sigma_{ij}^2(\mathbf{A}) = \sigma^2_{\max}=2 K \sigma_0^2$ for the utility differences $U_{i}-U_{j}$ and the correlations  $\varrho_i(\mathbf{A})=1/2$do not depend on the particular choice set $\mathbf{A}$.
Additionally, in the present case of indifference the preference probabilities are equal ($p_1=p_2=p_3=1/3$), and the intensity matrix amounts to $\bdgr{\Lambda }=  9(3\mathbf{I}_3-\mathbf{1}_3 \mathbf{1}_3^{\top})/(8\pi)$ for every choice set $\mathbf{A}$ and is, thus, constant within and across the orbits.
Hence, for each comparison depth $\mathbf{d}$ the uniform design $\bar{\xi}_{\mathbf{d}}$ on its orbit has the information matrix $\mathbf{M}( \bar{\xi}_{\mathbf{d}})= 9 \mathbf{I}_{K}/\pi$, which is independent of the orbit. Consequently any design $\bar{\xi}_{\mathbf{d}}$ is $D$-optimal as well as any convex combination thereof. This proves the following result.
\\[3mm]
\textbf{Theorem 4. } \textit{In the case of indifference $({\bdgr\beta}=\mathbf{0})$ every design, which is uniform on orbits with mean comparison depth $D=K$, is $D$-optimal under Model~I of independent utilities.}
\vspace{3mm}

Under the assumptions of Model~II the variances and correlations of the utility differences may vary with the orbits described by $\mathbf{d}$.
For a choice set  $\mathbf{A}$ with comparison depth $\mathbf{d}$ we get $\sigma_{ij}^2(\mathbf{A})= d_{ij}\sigma^2_{\max}/K$.
If additionally $d_{ij}>0$ for all $i$ and $j$ we obtain for the correlations
\[
\varrho_i(\mathbf{A})=(K-d_{j\ell})/\sqrt{ d_{ij} d_{i\ell}} \geq  0 \, .
\]
Consequently, the intensity matrix $\bdgr{\Lambda }$ does not vary for the choice sets within an orbit.
Then it can be seen that the mean intensity becomes
\[
\lambda_{\,\mathbf{d}}= \frac{1}{4\pi}\sum_{j=1}^{3}\frac{1+\varrho_{\,j}}{p_j} \, ,
\]
where the preference probabilities are given by
\[
p_j=\bdgr{\Phi }_{\varrho_j}(\mathbf{0})= \frac{1}{4} + \frac{\mathrm{arc}\,\sin(\varrho_{\,j})}{2\pi}  .
\]
It is worth-while mentioning that under the assumptions of Model~II the individual alternatives need not have equal preference probabilities even in the case of ``indifference'' ${\bdgr\beta}=\mathbf{0}$) due to the correlations between the utilities.

The situation of a choice set with two identical alternatives with comparison depth $\mathbf{d}=(K,K,0)$ can be covered by the paired comparison case of Section~3.

In Table~2 we present the preference probabilities and the normalized values $\sigma_{\max}^2\det(\mathbf{M}( \bar{\xi}_{\mathbf{d}}))^{1/K}$ of the criterion function together with the corresponding efficiencies
$\mbox{eff}\,(\bar{\xi}_{\mathbf{d}})=(\det\mathbf{M}( \bar{\xi}_{\mathbf{d}})/\det(\mathbf{M}( \bar{\xi}_{\mathbf{d}^*})) )^{1/K}=\lambda_{\mathbf{d}}/\lambda_{\mathbf{d}^*}$
for $K=2,...,7$ attributes and all possible comparison depths $\mathbf{d}$ with $d_{12}\geq d_{13}\geq d_{23}$.
For each number $K$ of attributes the optimal comparison depths $\mathbf{d}^*$ are highlighted in bold.
\begin{table}
\begin{center}
\begin{tabular}{ccccccccc}
$K$ & $d_{12}$ & $d_{13}$ & $d_{23}$ &\phantom{??} $p_1^*$ \phantom{??}& \phantom{??} $p_2^*$ \phantom{??} & \phantom{??} $p_3^*$ \phantom{??} & $\sigma_{\max}^2\det (\mathbf{M}( \bar{\xi}_{\mathbf{d}}))^{1/K}$ & $\mbox{eff}$ \\
 & & & & &  \multicolumn{2} {c} {($p_2^*+p_3^*$)}  & & \\
\hline
2 & 2 & 2 & 0 & 0.500 & \multicolumn{2} {c} {(0.500)} & 2.546 & 0.610 \\
\textbf{2} & \textbf{2} & \textbf{1} & \textbf{1} & \textbf{0.375} & \textbf{0.375} & \textbf{0.250} & \textbf{4.171} & \textbf{1.000} \\
\hline
3 & 3 & 3 & 0 & 0.500 & \multicolumn{2} {c} {(0.500)}& 2.546 & 0.593 \\
3 & 3 & 2 & 1 & 0.402 & 0.348 & 0.250 & 4.154 & 0.967 \\
\textbf{3} & \textbf{2} & \textbf{2} & \textbf{2} & \textbf{0.333} & \textbf{0.333} & \textbf{0.333} & \textbf{4.297} & \textbf{1.000} \\
\hline
4 & 4 & 4 & 0 & 0.500 & \multicolumn{2} {c} {(0.500)}& 2.546 & 0.595 \\
4 & 4 & 3 & 1 & 0.417 & 0.333 & 0.250 & 4.131 & 0.966 \\
4 & 4 & 2 & 2 & 0.375 & 0.375 & 0.250 & 4.171 & 0.975 \\
\textbf{4} & \textbf{3} & \textbf{3} & \textbf{2} & \textbf{0.366} & \textbf{0.317} & \textbf{0.317} & \textbf{4.278} & \textbf{1.000} \\
\hline
5 & 5 & 5 & 0 & 0.500 & \multicolumn{2} {c} {(0.500)}& 2.546 & 0.595 \\
5 & 5 & 4 & 1 & 0.426 & 0.324 & 0.250 & 4.111 & 0.960 \\
5 & 5 & 3 & 2 & 0.391 & 0.359 & 0.250 & 4.165 & 0.973 \\
5 & 4 & 4 & 2 & 0.385 & 0.308 & 0.308 & 4.249 & 0.992 \\
\textbf{5} & \textbf{4} & \textbf{3} & \textbf{3} & \textbf{0.348} & \textbf{0.348} & \textbf{0.304} & \textbf{4.282} & \textbf{1.000} \\
\hline
6 & 6 & 6 & 0 & 0.500 &\multicolumn{2} {c} {(0.500)}& 2.546 & 0.593 \\
6 & 6 & 5 & 1 & 0.433 & 0.317 & 0.250 & 4.094 & 0.953 \\
6 & 6 & 4 & 2 & 0.402 & 0.348 & 0.250 & 4.154 & 0.967 \\
6 & 6 & 3 & 3 & 0.375 & 0.375 & 0.250 & 4.171 & 0.971 \\
6 & 5 & 5 & 2 & 0.398 & 0.301 & 0.301 & 4.223 & 0.983 \\
6 & 5 & 4 & 3 & 0.367 & 0.336 & 0.297 & 4.267 & 0.993 \\
\textbf{6} & \textbf{4} & \textbf{4} & \textbf{4} & \textbf{0.333} & \textbf{0.333} & \textbf{0.333} & \textbf{4.297} & \textbf{1.000} \\
\hline
7 & 7 & 7 & 0 & 0.500 & \multicolumn{2} {c} {(0.500)} & 2.546 & 0.594 \\
7 & 7 & 6 & 1 & 0.438 & 0.312 & 0.250 & 4.079 & 0.951 \\
7 & 7 & 5 & 2 & 0.410 & 0.340 & 0.250 & 4.143 & 0.966 \\
7 & 7 & 4 & 3 & 0.386 & 0.364 & 0.250 & 4.168 & 0.972 \\
7 & 6 & 6 & 2 & 0.407 & 0.297 & 0.297 & 4.201 & 0.979 \\
7 & 6 & 5 & 3 & 0.380 & 0.328 & 0.292 & 4.249 & 0.990 \\
7 & 6 & 4 & 4 & 0.355 & 0.355 & 0.290 & 4.263 & 0.994 \\
\textbf{7} & \textbf{5} & \textbf{5} & \textbf{4} & \textbf{0.352} & \textbf{0.324} & \textbf{0.324} & \textbf{4.291} & \textbf{1.000} \\
\end{tabular}
\caption{Qualitative attributes: Characteristics of uniform designs $\bar{\xi}_{\mathbf{d}}$ for all comparison depths $\mathbf{d}$ with full profile under Model~II}
\end{center}
\end{table}

In the particular situation $\mathbf{d}=(K,K,0)$ the alternatives $2$ and $3$ are indistinguishable, and either of them may be chosen, if $U_2 = U_3 > U_1$, which occurs with probability $1/2$ as $U_1$ and $U_2$ are independent and identically distributed.
Then for the preference probabilities we have $p_1=1/2 = p_2 + p_3$, and the value of the normalized criterion function equals $\det (\mathbf{M}( \bar{\xi}_{\mathbf{d}}))^{1/K} = 8/(\pi  \sigma^2_{\max})$.

From Table~2 we can deduce for $K \leq 7$ that the maximal value of $\det(\mathbf{M}(\bar{\xi}_{\mathbf{d}}))$ is achieved for designs that are concentrated on those orbits, where the numbers of attributes, in which any two alternatives differ, are as balanced as possible.
It can be shown by convexity arguments that this statement holds true for all $K$, which are multiples of three, such that the optimal orbit is specified by $d_{ij} = 2K/3$ for all pairs of alternatives.
We conjecture that this result will be valid for any number of attributes $K$.

Note that the efficiencies of the choice sets with identical alternatives (comparison depth $\mathbf{d}=(K,K,0)$) are remarkably low.
%%%%%%%%%%%%%%%%%%%%%%%%%%%%%%%%%%%%%%%%%%%%%%%%%%%%%%%%%%%%%%%%%%%%%%%%%%%%%%%%%%%%%%%%%%%%%%%%%%%%%%%%%%
\subsection{One additional quantitative attribute}
For a model similar to that introduced by Kanninen (2002) we augment the above model with an additional continuous attribute $t$ as in Subsection~3.2.
For each alternative the set of attributes can be split into two components $\mathbf{a}_j=(\mathbf{x}_j,t_j)$, where $\mathbf{x}_j$ consists of the qualitative attributes and $t_j \in \mathbb{R}$.
For a choice set $\mathbf{A}=(\mathbf{a}_1,\mathbf{a}_2,\mathbf{a}_3)$ the marginal choice sets are denoted by $\mathbf{x}=(\mathbf{x}_1,\mathbf{x}_2,\mathbf{x}_3)$ and $\mathbf{t}=(t_1,t_2,t_3)$, respectively, and we can split the regression functions accordingly, $\mathbf{F}(\mathbf{x},\mathbf{t})=(\mathbf{F}_{1}(\mathbf{x})^{\top},\mathbf{F}_{2}(\mathbf{t}))^{\top}$, with marginal regression functions defined by $\mathbf{F}_{1}(\mathbf{x})=(\mathbf{f}(\mathbf{x_1}),\mathbf{f}(\mathbf{x_2}),\mathbf{f}(\mathbf{x_3}))$ and $\mathbf{F}_{2}(\mathbf{t})=(t_1,t_2,t_3)$.
The utilities $U_j(\mathbf{a}_j)$ in this two component model are generated from the part-worth utilities in the same way as in the paired comparison situation.

As before we assume that the first component consists of $K$ qualitative attributes with two levels each. By $\bar{\xi}_{1;\mathbf{d}}$ we denote a uniform marginal design on an orbit $\mathbf{d}=(d_{12},d_{13},d_{23})$, which involves the qualitative attributes only.

We start again with the standardized case, where $\bdgr{\beta}_1=\mathbf{0}$ and $\beta_2=1$.
There the intensity matrix $\bdgr{\Lambda}=\bdgr{\Lambda}_{\mathbf{d}}(\mathbf{t})$ depends only on the second component $\mathbf{t}=(t_1,t_2,t_3)$ and in addition on the scaling factors $\sigma_{ij}$ and the correlations $\varrho_i$, which may vary with the orbit~$\mathbf{d}$.

First we note that for a product type design $\bar{\xi}_{1_i \mathbf{d}} \otimes \xi_2$ with uniform marginal design $\bar{\xi}_{1;\mathbf{d}}$ on the orbit $\mathbf{d}=(d_{12},d_{13},d_{23})$ and arbitrary marginal design $\xi_2$ on the quantitative attribute the information matrix
\[
\mathbf{M}(\bar{\xi}_{1;\mathbf{d}}\otimes \xi_2)=\left(
\begin{tabular}{cc}
$4\left({\textstyle{\int}}\lambda_{2;\mathbf{d}}(\mathbf{t}) \xi_2(\mathrm{d}\mathbf{t})\right) \mathbf{I}_K $& $\mathbf{0}$\\
$\mathbf{0}$  & ${\textstyle{\int}} m_{\mathbf{d}}(\mathbf{t})\, \xi_2(\mathrm{d}\mathbf{t})$
\end{tabular}
\right)
\]
is diagonal, where $m_{\mathbf{d}}(\mathbf{t})=\mathbf{t} \bdgr{\Lambda }_{\mathbf{d}}(\mathbf{t}) \mathbf{t}^{\top}$, $\lambda_{2; \mathbf{d}}(\mathbf{t})= \frac{1}{K}\sum_{j=1}^{3}(K-d_{i\ell})\lambda_{\mathbf{d},jj}(\mathbf{t})$ is the mean intensity on the orbit $\mathbf{d}$ and $\lambda_{\mathbf{d},jj}$ is the $j$th diagonal element of $\bdgr{\Lambda}_{\mathbf{d}}$.
Then the determinant of the information matrix $\mathbf{M}(\bar{\xi}_{1;\mathbf{d}}\otimes \xi_2)$ becomes
\[
\det(\mathbf{M}(\bar{\xi}_{1;\mathbf{d}}\otimes \xi_2))=\left( 4 {\textstyle{\int}}\lambda_{2 ;\mathbf{d}}(\mathbf{t})\, \xi_2(\mathrm{d}\mathbf{t})\right)^K\, {\textstyle{\int}} m_{\mathbf{d}}(\mathbf{t}) \xi_2(\mathrm{d}\mathbf{t}) \, .
\]

Under the assumptions of Model~I all part-worth utilities for the first component are assumed to be independent. Then the variances of the utility differences are again $\sigma_{ij}^2(\mathbf{A})=2\,( K\sigma_0^2+\sigma_t^2)=\sigma^2_0$.
We conjecture that the determinant of the information matrix $\mathbf{M}(\bar{\xi}_{1;\mathbf{d}}\otimes \xi_2)$ will be maximized by a marginal one point design $\xi_2=\delta _{\mathbf{t}}$ for a suitable optimal setting $\mathbf{t}=\mathbf{t}^*$ of the second component.
Numerically the maximization of the determinant $\det(\mathbf{M}(\bar{\xi}_{1;\mathbf{d}}\otimes \delta _{\mathbf{z}}))$ was carried out with respect to $\mathbf{t}$ for $K\leq 7$ qualitative attributes and for all possible comparison depths $\mathbf{d}$ with $d_{12}\geq d_{13}\geq d_{23}$ of full profile ($D=K$).
There we used the standardized version $\mathbf{z}=(z_1,z_2,z_3)$ with $z_j=t_j-t_3$ for the second component, as the choice probabilities are invariant with respect to a shift of location.
Because of $z_3=0$ then only $z_1$ and $z_2$ have to be optimized.

In Table~3 we present the optimal values $z_1^*$ and $z_2^*$ for the quantitative attribute, the corresponding choice probabilities $p_j^*$ and the normalized values $\sigma_{\max}^2\det(\mathbf{M}( \bar{\xi}_{\mathbf{d}}))^{1/(K+1)}$ of the criterion function together with their associated efficiencies
$\mbox{eff}\,(\bar{\xi}_{\mathbf{d}})=(\det\mathbf{M}( \bar{\xi}_{\mathbf{d}})/\det(\mathbf{M}( \bar{\xi}_{\mathbf{d}^*})) )^{1/(K+1)}$
for $K\leq 7$ attributes and all possible comparison depths $\mathbf{d}$ with $d_{12}\geq d_{13}\geq d_{23}$. The optimal comparison depths $\mathbf{d}^*$ are highlighted in bold for each $K$.
In all cases the maximal value for the determinant is achieved for the design concentrated on the orbits with two identical alternatives.
This coincides with the findings in the logistic case observed in Grasshoff et al.\ (2013).
\begin{table}
\begin{center}
\scriptsize
\begin{tabular}{ccccccccccc}
$K$ & $d_{12}$ & $d_{13}$ & $d_{23}$ & \phantom{??} $z_1^*$ \phantom{??} & \phantom{??} $z_2^*$ \phantom{??}& \phantom{??} $p_1^*$ \phantom{??}&
\phantom{??} $p_2^*$ \phantom{??} & \phantom{??} $p_3^*$ \phantom{??} & $\sigma_{\max}^2\det (M)^{1/(K+1)}$ & $\mbox{eff}$\\ \hline
\textbf{1} & \textbf{1} & \textbf{1} & \textbf{0} & \textbf{1.26} & \textbf{0.00} & \textbf{0.827} & \textbf{0.087} & \textbf{0.087} & \textbf{1.344} & \textbf{1.000} \\
\hline
\textbf{2} & \textbf{2} & \textbf{2} & \textbf{0} & \textbf{1.07} & \textbf{0.00} & \textbf{0.769} & \textbf{0.116} & \textbf{0.116} & \textbf{1.609} & \textbf{1.000} \\
2 & 2 & 1 & 1 & 1.33 & 0.55 & 0.741 & 0.199 & 0.060 & 1.504 & 0.935 \\
\hline
\textbf{3} & \textbf{3} & \textbf{3} & \textbf{0} & \textbf{0.96} & \textbf{0.00} & \textbf{0.731} & \textbf{0.134} & \textbf{0.134} & \textbf{1.801} & \textbf{1.000} \\
3 & 3 & 2 & 1 & 1.21 & 0.55 & 0.698 & 0.231 & 0.071 & 1.720 & 0.955 \\
3 & 2 & 2 & 2 & 0.88 & 0.00 & 0.702 & 0.149 & 0.149 & 1.547 & 0.859 \\
\hline
\textbf{4} & \textbf{4} & \textbf{4} & \textbf{0} & \textbf{0.88} & \textbf{0.00} & \textbf{0.702} & \textbf{0.149} & \textbf{0.149} & \textbf{1.947} & \textbf{1.000} \\
4 & 4 & 3 & 1 & 1.12 & 0.54 & 0.667 & 0.252 & 0.081 & 1.881 & 0.966 \\
4 & 4 & 2 & 2 & 1.19 & 0.75 & 0.632 & 0.305 & 0.063 & 1.848 & 0.949 \\
4 & 3 & 3 & 2 & 0.82 & 0.00 & 0.679 & 0.161 & 0.161 & 1.740 & 0.894 \\
\hline
\textbf{5} & \textbf{5} & \textbf{5} & \textbf{0} & \textbf{0.83} & \textbf{0.00} & \textbf{0.681} & \textbf{0.159} & \textbf{0.159} & \textbf{2.060} & \textbf{1.000} \\
5 & 5 & 4 & 1 & 1.06 & 0.54 & 0.643 & 0.269 & 0.088 & 2.006 & 0.973 \\
5 & 5 & 3 & 2 & 1.12 & 0.75 & 0.603 & 0.327 & 0.069 & 1.978 & 0.960 \\
5 & 4 & 4 & 2 & 0.77 & 0.00 & 0.659 & 0.170 & 0.170 & 1.886 & 0.915 \\
5 & 4 & 3 & 3 & 0.92 & 0.39 & 0.628 & 0.256 & 0.116 & 1.823 & 0.885 \\
\hline
\textbf{6} & \textbf{6} & \textbf{6} & \textbf{0} & \textbf{0.78} & \textbf{0.00} & \textbf{0.663} & \textbf{0.168} & \textbf{0.168} & \textbf{2.152} & \textbf{1.000} \\
6 & 6 & 5 & 1 & 1.01 & 0.53 & 0.626 & 0.280 & 0.094 & 2.105 & 0.978 \\
6 & 6 & 4 & 2 & 1.07 & 0.74 & 0.586 & 0.340 & 0.074 & 2.081 & 0.967 \\
6 & 6 & 3 & 3 & 1.05 & 0.93 & 0.514 & 0.422 & 0.064 & 2.069 & 0.962 \\
6 & 5 & 5 & 2 & 0.73 & 0.00 & 0.643 & 0.178 & 0.178 & 2.001 & 0.930 \\
6 & 5 & 4 & 3 & 0.88 & 0.38 & 0.614 & 0.263 & 0.122 & 1.947 & 0.905 \\
6 & 4 & 4 & 4 & 0.67 & 0.00 & 0.618 & 0.191 & 0.191 & 1.857 & 0.863 \\
\hline
\textbf{7} & \textbf{7} & \textbf{7} & \textbf{0} & \textbf{0.75} & \textbf{0.00} & \textbf{0.651} & \textbf{0.174} & \textbf{0.174} & \textbf{2.227} & \textbf{1.000} \\
7 & 7 & 6 & 1 & 0.97 & 0.52 & 0.612 & 0.288 & 0.100 & 2.185 & 0.981 \\
7 & 7 & 5 & 2 & 1.02 & 0.72 & 0.572 & 0.348 & 0.080 & 2.165 & 0.972 \\
7 & 7 & 4 & 3 & 1.02 & 0.87 & 0.522 & 0.408 & 0.070 & 2.154 & 0.967 \\
7 & 6 & 6 & 2 & 0.70 & 0.00 & 0.631 & 0.185 & 0.185 & 2.095 & 0.941 \\
7 & 6 & 5 & 3 & 0.85 & 0.37 & 0.605 & 0.268 & 0.127 & 2.047 & 0.919 \\
7 & 6 & 4 & 4 & 0.90 & 0.62 & 0.553 & 0.347 & 0.100 & 2.023 & 0.908 \\
7 & 5 & 5 & 4 & 0.65 & 0.00 & 0.610 & 0.195 & 0.195 & 1.968 & 0.884 \\
\end{tabular}
\caption{One additional quantitative attribute: Optimal values for $\mathbf{z} = (z_1 , z_2,0)$, optimal choice probabilities and design characteristics for orbits $\mathbf{d}$ under Model~I, $K \leq 7 $.}
\end{center}
\end{table}

Under the assumptions of Model~II for the first component of qualitative attributes the variances of the utility differences between the $i$th and $j$th alternative and the corresponding correlations are given by
\[
\sigma_{ij}^2(\mathbf{A})=(d_{ij}\sigma_{\max}^2+2 (K-d_{ij})\sigma_t^2)/K
\]
and
\[
\varrho_j(\mathbf{A})=((K-d_{i\ell}) \sigma_{\max}^2-(K-2d_{i\ell}) \sigma_t^2)/(K \sigma_{ij} \sigma_{j\ell})
\]
for any choice set $\mathbf{A}$ of comparison depth $\mathbf{d}$.
%With this notation of subsection $3.2$ we get
%\[
%\sigma_{ij}(\mathbf{A})^2= \left( (1-\varepsilon )\frac{d_{ij}}{K} + \varepsilon \right) \, \sigma^2_0
%\]
Also here the case $\sigma_t^2 =0$ represents a sharp decision concerning the quantitative variable $\mathbf{t}$.
The corresponding numerical results for such a sharp decision are exhibited in Table~4.
There we present the optimal values for $z_1^*$ and $z_2^*$ for the quantitative attribute, the corresponding choice probabilities $p_j^*$ and the normalized values $\sigma_{\max}^2\det(\mathbf{M}( \bar{\xi}_{\mathbf{d}}))^{1/(K+1)}$ of the criterion function together with the associated efficiencies
$\mbox{eff}\,(\bar{\xi}_{\mathbf{d}})=(\det\mathbf{M}( \bar{\xi}_{\mathbf{d}})/\det(\mathbf{M}( \bar{\xi}_{\mathbf{d}^*})) )^{1/(K+1)}$
for $K\leq 7$ attributes and all possible comparison depths $\mathbf{d}$ with $d_{12}\geq d_{13}\geq d_{23}$.
The optimal comparison depths $\mathbf{d}^*$ are highlighted in bold for each $K$.
In all cases ($K\geq 2$) the maximal value for the determinant is achieved for the designs concentrated on the orbits with two alternatives, which differ only in one qualitative attribute.
However, if the decision is not sharp ($\sigma_t^2>0$), we found out numerically that other comparison depths may become optimal, where the alternatives differ in more than one qualitative attribute.
\begin{table}
\scriptsize
\begin{center}
\begin{tabular}{ccccccccccc}
$K$ & $d_{12}$ & $d_{13}$ & $d_{23}$ & \phantom{??} $z_1^*$ \phantom{??} & \phantom{??} $z_2^*$ \phantom{??}& \phantom{??} $p_1^*$ \phantom{??}&
\phantom{??} $p_2^*$ \phantom{??} & \phantom{??} $p_3^*$ \phantom{??} & $\sigma_{\max}^2\det (M)^{1/(K+1)}$ & $\mbox{eff}$\\
& & & & & & &  \multicolumn{2} {c} {($p_2^*+p_3^*$)}  & & \\
\hline
\textbf{1} & \textbf{1} & \textbf{1} & \textbf{0} & \textbf{-1.14} &  \textbf{0.00} & \textbf{0.127} & \multicolumn{2}{c}{\textbf{(0.873)}}& \textbf{0.891} & \textbf{1.000}\\
\hline
2 & 2 & 2 & 0 & -0.94 &  0.00 & 0.174 & \multicolumn{2}{c} {(0.826)}& 1.109 & 0.540\\
\textbf{2} & \textbf{2} & \textbf{1} & \textbf{1} & \textbf{-0.72} &  \textbf{0.00} & \textbf{0.142} & \textbf{0.142} & \textbf{0.715} & \textbf{2.054} & \textbf{1.000}\\
\hline
3 & 3 & 3 & 0 & -0.82 &  0.00 & 0.206 & \multicolumn{2}{c} {(0.794)} & 1.272 & 0.546\\
\textbf{3} & \textbf{3} & \textbf{2} & \textbf{1} & \textbf{-0.77} & \textbf{-0.49} & \textbf{0.159} & \textbf{0.178} & \textbf{0.663} & \textbf{2.328} & \textbf{1.000}\\
3 & 2 & 2 & 2 & -0.72 & -0.72 & 0.149 & 0.149 & 0.702 & 2.097 & 0.901\\
\hline
4 & 4 & 4 & 0 & -0.73 &  0.00 & 0.233 & \multicolumn{2}{c} {(0.767)} & 1.398 & 0.551\\
\textbf{4} & \textbf{4} & \textbf{3} & \textbf{1} & \textbf{-0.78} & \textbf{-0.37} & \textbf{0.168} & \textbf{0.203} & \textbf{0.629} & \textbf{2.537} & \textbf{1.000}\\
4 & 4 & 2 & 2 & -0.58 & -0.58 & 0.185 & 0.185 & 0.630 & 2.536 & 1.000\\
4 & 3 & 3 & 2 & -0.74 & -0.52 & 0.158 & 0.185 & 0.657 & 2.364 & 0.932\\
\hline
5 & 5 & 5 & 0 & -0.67 &  0.00 & 0.251 & \multicolumn{2}{c} {(0.749)} & 1.500 & 0.555\\
\textbf{5} & \textbf{5} & \textbf{4} & \textbf{1} & \textbf{-0.77} & \textbf{-0.29} & \textbf{0.178} & \textbf{0.225} & \textbf{0.597} & \textbf{2.702} & \textbf{1.000}\\
5 & 5 & 3 & 2 & -0.62 & -0.46 & 0.189 & 0.206 & 0.604 & 2.701 & 1.000\\
5 & 4 & 4 & 2 & -0.75 & -0.39 & 0.162 & 0.218 & 0.620 & 2.566 & 0.950\\
5 & 4 & 3 & 3 & -0.57 & -0.58 & 0.188 & 0.184 & 0.628 & 2.572 & 0.952\\
\hline
6 & 6 & 6 & 0 & -0.62 &  0.00 & 0.268 & \multicolumn{2}{c} {(0.732)} & 1.583 & 0.558\\
\textbf{6} & \textbf{6} & \textbf{5} & \textbf{1} & \textbf{-0.77} & \textbf{-0.23} & \textbf{0.181} & \textbf{0.247} & \textbf{0.571} & \textbf{2.837} & \textbf{1.000}\\
6 & 6 & 4 & 2 & -0.63 & -0.38 & 0.197 & 0.222 & 0.581 & 2.835 & 0.999\\
6 & 6 & 3 & 3 & -0.51 & -0.51 & 0.208 & 0.208 & 0.585 & 2.835 & 0.999\\
6 & 5 & 5 & 2 & -0.75 & -0.30 & 0.166 & 0.245 & 0.588 & 2.727 & 0.961\\
6 & 5 & 4 & 3 & -0.60 & -0.46 & 0.189 & 0.210 & 0.601 & 2.733 & 0.963\\
6 & 4 & 4 & 4 & -0.55 & -0.55 & 0.620 & 0.190 & 0.190 & 2.629 & 0.927\\
\hline
7 & 7 & 7 & 0 & -0.58 &  0.00 & 0.281 & \multicolumn{2}{c} {(0.719)} & 1.652 & 0.560\\
\textbf{7} & \textbf{7} & \textbf{6} & \textbf{1} & \textbf{-0.76} & \textbf{-0.19} & \textbf{0.187} & \textbf{0.263} & \textbf{0.550} & \textbf{2.948} & \textbf{1.000}\\
7 & 7 & 5 & 2 & -0.64 & -0.32 & 0.200 & 0.237 & 0.563 & 2.946 & 0.999\\
7 & 7 & 4 & 3 & -0.54 & -0.43 & 0.209 & 0.223 & 0.568 & 2.945 & 0.999\\
7 & 6 & 6 & 2 & -0.75 & -0.23 & 0.168 & 0.272 & 0.559 & 2.858 & 0.969\\
7 & 6 & 5 & 3 & -0.62 & -0.38 & 0.190 & 0.229 & 0.581 & 2.863 & 0.971\\
7 & 6 & 4 & 4 & -0.50 & -0.50 & 0.208 & 0.208 & 0.583 & 2.864 & 0.972\\
7 & 5 & 5 & 4 & -0.58 & -0.44 & 0.187 & 0.218 & 0.595 & 2.777 & 0.942\\
\end{tabular}
\end{center}
\caption{One additional quantitative attribute:  Optimal values for $\mathbf{z} = (z_1 , z_2,0)$, optimal choice probabilities and design characteristics for orbits $\mathbf{d}$ under Model~II with sharp decision, $K \leq 7 $}
\end{table}

Note that similar to Subsection~4.1 in the case of sharp decisions the alternatives $\mathbf{a}_2$ and $\mathbf{a}_3$ are indistinguishable ($U_2 = U_3$) for choice sets with comparison depth $\mathbf{d} = (K, K, 0)$, if $z^*_2 = 0$, i.\,e.\ if the quantitative attribute is set to the same level for both alternatives.
If in this case $z_2^* > 0$ (or $z_2^* < 0$) there will be a strict preference of alternative $\mathbf{a}_2$ over $\mathbf{a}_3$ ($\mathbf{a}_3$ over $\mathbf{a}_2$, respectively) such that essentially we end up in a paired comparison situation for the pair  $(\mathbf{a}_1 ,  \mathbf{a}_2)$ (resp.\ $(\mathbf{a}_1 ,  \mathbf{a}_3)$) of alternatives with the same value for the information matrix as specified in Table~4.
This may explain, why in this situation the efficiencies of choice sets with comparison depth $\mathbf{d}=(K,K,0)$ are so low such that the counter-intuitive result of Model~I does not occur.
\section{Discussion}
This paper provides an important extension of previous developments of optimal designs for discrete choice models (for an overview, see Gro{\ss}mann and Schwabe, 2015). The designs for multinomial discrete choice models derived so far do not seem appropriate for many practical purposes due to the IIA property. However, merely changing the link function to a probit one does not alleviate the problem as shown above.
Probit models allow for introducing dependencies between the part and, thus, the overall utilities so that many choice situations can be modelled more appropriately.
We developed locally D-optimal designs assuming choice sets all consisting of either two or three options. According to a further assumption which is also typical of many choice experiments, each respondent is faced with the same sets of choices. The derived D-optimal designs for the case of indifference may not be very important for many choice situations in practice. However, they are the starting points for deriving such designs for the more general case of any parameter values, as D-optimal designs for linear models and locally D-optimal multinomial models coincide. Thus, the concept of canonical transformation (Sitter \& Torsney, 2007) could be applied.
Further developments concerning the designs for discrete choice models based on probit regression with dependent utilities should consider more than two levels for the discrete attributes and, furthermore, be extended to several quantitative attributes (see Kannninen, 2002). It would also be interesting to use further optimality criteria instead of D-optimality, such as IMSE-optimality.

\section*{Appendix: Proofs}
In order to apply the constructions of Gra{\ss}hoff et al. (2007) to the present probit paired comparison situation of Section~3 we make use of the following auxiliary results.
We start with some useful inequalities for the normal distribution.\\[2ex]
%%%%%%%%%%%%%%%%%%%%%%%%%%%%%%%%%%%%%%%%%%%%%%%%%%%%%%%%%%%%%%%%%%%%%%%%%%%%%%%%%%%%5
\textbf{Lemma 1.}\\[1ex]
\textit{
a) $ 1 - \Phi_0 (z) \geq \left( 1 - \frac{ z^2 + 7}{ 8z^2 + 12} \right)  \frac{ 1}{ z} \, \varphi_0 (z)\quad $ for $z \geq 1$,\\[1ex]
b) $\Phi_0 (z) - \frac{ 1}{ 2} \leq \frac{ 1}{ \sqrt{2 \pi}}  \left( z - \frac{ 1}{ 6} \, z^3 + \frac{ 1}{ 40} \, z^5 \right)\quad $ for $0 \leq z \leq 1$.}\\[2ex]
\textbf{Proof.}
Assertion a) follows along the lines for standard lower bounds of the tail probability (see e.\,g.\ G\"anssler and Stute, 1977, p.\ 105):\\[1ex]
Let $\tau (z) = - \left( 1- \frac{ z^2 + 7}{ 8z^2 + 12} \right) \, \frac{ 1}{ z} \varphi_0 (z)$. Then for the derivative $\tau^{\prime}$ it holds
$$
\tau^{\prime} (z) = \varphi_0 (z)  \left( 1 - \frac{ 2z^6 + 3z^4 + 12 z^2 - 15}{ z^2, (4z^2 + 6)^2} \right) \leq \varphi_0 (z)
$$
for all $z \geq 1$.
Hence, for the tail probability
$$
1 - \Phi_0 (z) = \int\limits^{\infty}_z \varphi_0 (x) \mathrm{d}x \geq \int\limits^{\infty}_z  \tau^{\prime} (x) \mathrm{d}x = - \tau (z)
$$
as $\tau (z) \to 0$ for $z \to \infty$, which proves a).

Assertion b) follows by Taylor expansion up to terms of order five at $z_0 = 0$, as the even coefficients are vanishing and the odd coefficients are alternating and decreasing.
 \hfill $\Box$

Next we derive an auxiliary property of the function $h(z) = 1 / \lambda_2 (z)$.\\[2ex]
%%%%%%%%%%%%%%%%%%%%%%%%%%%%%%%%%%%%%%%%%%%%%%%%%%%%%%%%%%%%%%%%%%%%%%%%%%%%%%%%%%%%%%%%%%%%%%%%%%%%%%%%%%%%%%%%%%%%%%%%%%
\textbf{Lemma 2.}
\textit{
Let $h(z) = \Phi_0 (z)  (1 - \Phi_0 (z))/\varphi_0 (z)^2$. Then the third derivative $h^{\prime \prime \prime}$ satisfies $h^{\prime \prime \prime} (z) > 0 $ for all $z > 0$.}\\[2ex]
\textbf{Proof.}
First note that
$$
h^{\prime \prime \prime} (z) = (8 z^3 + 12 z) \frac{ \Phi_0 (z) (1-\Phi_0 (z))}{ \varphi_0 (z)^2} - (14 z^2 + 10)  \frac{ \Phi_0 (z) -  1/ 2}{ \varphi_0 (z)} - 6 z
$$
and, hence, $h^{\prime\prime\prime}(0)=0$ and
\begin{eqnarray*}
\varphi_0 (z)^2  h^{\prime \prime \prime}  (z)&  = & 2z^3 + 3z - (8z^3 + 12z)  ( \Phi_0 (z) - 1/2)^2\\[0.5ex]
                                                     & & \mbox{} -  (14 z^2 + 10) ( \Phi_0 (z) - 1/2)  \varphi_0 (z) - 6z \varphi_0 (z)^2 \, .
\end{eqnarray*}
For $z \leq 1$ by Lemma 1 b) we have $(\Phi_0 (z) - 1/2)^2 \leq \frac{ 1}{ 2 \pi} \, \left( z^2 - \frac{ 1}{ 3} z^4 + \frac{ 7}{ 90} \, z^6 \right)\,.$
Using this, $e^{-x} \leq 1 - x + \frac{ 1}{ 2} \, x^2$ for $0 \leq x \leq 1$ applied to $x = z^2$ and $x = \frac{ 1}{ 2} \, z^2$, respectively, and Lemma 1 b) we obtain
\begin{eqnarray*}
2 \pi \varphi_0 (z)^2 h^{\prime \prime \prime}  (z) & \geq & 6 \pi z + 4 \pi z^3\\[0.5ex]
                                                         & &  \mbox{} -  (8z^3 + 12 z)  \left( z^2 - \frac{ 1}{ 3} \, z^4 + \frac{ 7}{ 90} z^6 \right)\\[0.5ex]
                                                         & & \mbox{} -  (14 z ^2 + 10)  \left( z - \frac{ 1}{ 6}  z^3 + \frac{ 1}{ 40}  z^5 \right) \, \left( 1- \frac{ 1}{ 2}  z^2 + \frac{ 1}{ 8}
                                                                        z^4 \right)\\[0.5ex]
                                                         & &  \mbox{} -   6z \left( 1 - z^2 + \frac{ 1}{ 2}  z^4 \right)\\[0.5ex]
                                                         &  = & (6 \pi - 16) z + \left( 4 \pi - \frac{ 40}{ 3} \right) z^3 - \frac{ 6}{ 5}  z^7 - \frac{ 269}{ 1440}  z^9 - \frac{ 7}{ 160}  z^{11}\\[0.5ex]
                                                         & \geq & 2.8 z - 0.8 z^3 - 1.2 z^7 - 0.2 z^9 - 0.1 z^{11}\\[0.5ex]
                                                         & \geq & 0.5 z \; ,
\end{eqnarray*}
which proves the assertion for $z \leq 1$.

For $z > 1$ we use the identity
\begin{eqnarray*}
\frac{ \varphi_0 (z)^2  h^{\prime \prime \prime} (z)}{ 2 z^3 + 3z} &  = & 1 - \left( 1-2 \left( ( 1 - \Phi_0 (z)) - \left( 1 - \frac{ z^2 + 7}{ 8 z^2 + 12} \right) \, \frac{ 1}{ z}  \varphi_0 (z) \right) \right)^2\\[0.5ex]
                                                                          & & \mbox{} +   \frac{ (z^2 - 1)^2 + 24}{ (4z^2 + 6)^2  z^2} \, \varphi_0 (z)^2 \; .
\end{eqnarray*}
By Lemma~1 a) the squared term is bounded by one for all $z \geq 1$, while the last expression is positive, which establishes the result.
\hfill $\Box$

%As a consequence we obtain for the auxiliary criterion $\Psi_2$ that the optimal design has minimal support.\\[2ex]
%%%%%%%%%%%%%%%%%%%%%%%%%%%%%%%%%%%%%%%%%%%%%%%%%%%%%%%%%%%%%%%%%%%%%%%%%%%%%%%%%
\textbf{Lemma 3.} \textit{Let $z^*>0$ be the unique maximum of the function $\lambda_2(z)^p \, z^2$.\\
Every design $\zeta^*$ which is concentrated on $\{-z^* \, , \; z^*\}$ maximizes the criterion
\[
 \Psi_2 (\zeta) = {\textstyle{\int}} z^2 \lambda_2 (z) \,  \zeta (\mathrm{d}z)  ( {\textstyle{\int}} \lambda_2 \mathrm{d}  \zeta )^{p-1} \, .
\]
}
\\
\textbf{Proof.}
The proof is similar to the situation of logistic response considered in Gra{\ss}hoff et al.\ (2007), Lemma 1, and uses an idea of Biedermann et al.\ (2006):

Let $\zeta^*$ be a $\Psi_2$-optimal design. Denote by $m^*_j = \int z^j \lambda_2 (z)  \zeta^* (\mathrm{d}z)$, $ j = 0,2$, the corresponding weighted moments involved in $\Psi_2$.
The equivalent criterion $\ln \Psi_2$ is concave and its directional derivative at $\zeta^*$ in the direction of the one point design in $z$ is
$$
\psi_2  (z) = \lambda_2 (z)  (z^2 /m^*_2 + (p-1) / m^*_0) - p \, .
$$
By the general equivalence theorem (see Silvey, 1980) the inequality $\psi_2 (z) \leq 0$ is satisfied for all $z$, and its maximum $\psi_2 (z) = 0$ is attained for $z$ in the support of $\zeta^*$.
Denote further by $h(z) = 1/\lambda_2 (z)$ the inverse intensity function. The above condition can then be rewritten as
$$
g(z) = h (z) - \frac{1}{p m^*_2}z^2 - \frac{p-1}{p m^*_0} \geq 0
$$
for all $z$, and equality holds for $z$ in the support of $\zeta^*$.
Note that $g$ is symmetric, $g(z)$ tends to infinity for $z \to \infty$, and the third derivative $g^{\prime \prime \prime} = h^{\prime \prime \prime}$ has only one root, $g^{\prime \prime \prime} (0) = 0$, according to Lemma 2.
As a consequence $g$ may have, at most, one local minimum $z_0 > 0$, say.
Thus, the optimal design $\zeta^*$ is concentrated on $\{ - z_0 , z_0 \}$ and, hence, $\Psi_2 (\zeta^*) = z_0^2  \lambda_2  (z_0)^p$, which is maximized by
$z_0 = z^*$.
\hfill $\Box$
\vspace{3mm}

For $K=1$ the information matrix of the paired comparison model can be identified with that of a standard probit model with one continuous explanatory variable.
In this situation, as a by-product, Lemma~3 gives an analytical proof for the corresponding result of minimal support established numerically in Biedermann et al.\ (2006).
\vspace{3mm}

\noindent
\textbf{Proof of Theorem~1. }
Because only the difference $t_1 - t_2$ is involved in the intensity, we consider $z=Z(\mathbf{t}) = t_1 - t_2$.
Let further $\delta_{\mathbf{t}}$ be the one-point design in $\mathbf{t}$.
Then the design $\delta^{Z}_{\mathbf{t}}$ induced by $Z$ is the one-point design $\delta_z$ in $z=Z(\mathbf{t})$.
By Lemma~3 the design $\delta_{z^*}$ maximizes $\Psi_2 (\zeta_2) = \int  z^2 \lambda_2 (z) \zeta_2 (\mathrm{d}z) ( \int  \lambda_2  \mathrm{d} \zeta_2)^K$.
As has been mentioned in Subsection~3.1 the uniform design $\bar{\xi}_K$ is $D$-optimal in the marginal model associated with the first component $\mathbf{x}$.
Due to the orthogonality property $\int \mathbf{F}_1 \mathrm{d} \bar{\xi}_K = \mathbf{0}$ of the uniform design $\bar{\xi}_K$ Theorem 2 in Gra{\ss}hoff et al.\ (2007) applies, which establishes that for $\mathbf{t}^*$ such that $Z(\mathbf{t}^*)=z^+$ the product type design
$\bar{\xi}_K \otimes \delta_{\mathbf{t}^*}$ is $D$-optimal for the probit paired comparison model with independent utilities.
\hfill $\Box$
\vspace{3mm}

The following result establishes that every design is dominated by a product type design for the model considered in Subsection~3.2.
\\[2ex]
%%%%%%%%%%%%%%%%%%%%%%%%%%%%%%%%%%%%%%%%%%%%%%%%%%%%%%%%%%%%%%%%%%%%%%%%%%%%%%%%%%%%%%%%%%%%%%%%%%%%%%%%%%%%%%%%%%%
\textbf{Lemma 4}. \textit{Under the assumptions of Subsection~3.2 tor every design $\xi$ there exists a marginal design $\tilde{\xi}_2$ such that
\[
\det (  \mathbf{M}(\xi) ) \leq \det( \mathbf{M}(\bar{\xi}_K \otimes \tilde{\xi}_2)  ) \, .
\]}
\\
\textbf{Proof.}
Let $\xi_1$ be the marginal design of $\xi$ on the first component and denote by $w_d$ the weight of $\xi_1$ on the orbit of comparison depth $d$.
The corresponding symmetrized design $\bar{\xi}$ with respect to permutations of the levels and attributes can be written as a
weighted sum $\bar{\xi}=\sum_{d=1}^K w_d \, \bar{\xi_d}\otimes \xi_{2;d}$ of designs $\bar{\xi}_d \otimes \xi_{2;d}$ concentrated on the orbits induced by the comparison depth $d$.
Here $\xi_{2;d}$ denotes the conditional marginal distribution of $\xi$ for the second component, conditionally on the orbit of comparison depth $d$.
Due to the invariance of the $D$-criterion the design $\xi$ is dominated by $\bar{\xi}$, i.\,e.\ $\det (\mathbf{M} (\xi))\leq \det( \mathbf{M} (\bar{\xi}))$
(see e.\,g.\ Schwabe, 1996, section~3.2).

Denote by $\sigma^2 (d)$
% $ = (1 - \varepsilon) \, \frac{ d}{ K} + \varepsilon$
the variance associated with comparison
depth $d$. The information matrix
\begin{eqnarray*}
\mathbf{M} (\bar{\xi}_{d}\otimes \xi_2 ) & = &
\left(
\begin{tabular}{cc}
$ c_1 (d , \xi_2 ) \, \mathbf{I}_{K}\otimes \mathbf{M}^* $& $\mathbf{0}$\\
$\mathbf{0}$ & $c_2(d , \xi_2)$
\end{tabular}
\right)\\
\end{eqnarray*}
of a product type design $\bar{\xi}_d\otimes \xi_2$ is block diagonal with coefficients
\[
c_1(d, \xi_2 ) \leq \int \lambda_2 ( z/ \sigma (d)) \xi_2^{Z} (\mathrm{d}z)
\]
with equality for $d=K$ and
\[
c_2 (d, \xi_2 ) = \int (z/\sigma (d))^2 \lambda_2(z/\sigma (d)) \xi_2^{Z} (\mathrm{d}z) \, ,
\]
where $Z (\mathbf{t}) = t_1 - t_2$ and $\xi_2^{Z}$ is the image of $\xi_2$ under $Z$ as in the proof of Theorem~1.

Substitute $\tilde{z} = z / \sigma (d)$ and let $\tilde{\xi}_{2;d}$ be the image of $\xi_{2;d}$ under this transformation.
Then we obtain
$ c_1(d, \xi_{2;d}) \leq c_1 (K , \, \tilde{\xi}_{2;d})$ and
$ c_2(d, \xi_{2; d}) = c_2 \, (K , \, \tilde{\xi}_{2;d})$. This implies
$\mathbf{M} (\bar{\xi}_{d}\otimes \xi_{2;d}) \leq \mathbf{M} (\bar{\xi}_{K}\otimes \tilde{\xi}_{2;d})$ and, consequently,
$$
\mathbf{M} (\bar{\xi}) \leq  \sum_{d=1}^K w_d \, \mathbf{M} (\bar{\xi}_{K}\otimes \tilde{\xi}_{2;d}) = \mathbf{M} (\bar{\xi}_K\otimes \tilde{\xi}_2) \, ,
$$
where  $\tilde{\xi}_2$ is defined by $\tilde{\xi}_2=\sum_{d=1}^{K} w_d\, \tilde{\xi}_{2;d} $. This completes the proof.
\hfill $\Box$\\[2ex]
%%%%%%%%%%%%%%%%%%%%%%%%%%%%%%%%%%%%%%%%%%%%%%%%%%%%%

\textbf{Proof of Theorem~2. } Let again $Z (\mathbf{t}) = t_1 - t_2$. Since $\det ( \bdgr{M}(\bar{\xi}_K\otimes \xi_2) )=\Psi_2(\xi^{Z}_2)$ and $\delta_{z^*}$ maximizes $\Psi_2$,
the result follows directly from Lemma~4.
\hfill $\Box$

\vspace{5mm}
\noindent
{\bf Acknowledgments:}  Research supported by the Deutsche Forschungsgemeinschaft (DFG) under grant HO 1286/6. Part of the work was done, while the last author was visiting the International Newton Institute in Cambridge. The authors wish to express their thanks to Robert Offinger for providing the proof of Lemma~1.

\vspace{5mm}
\noindent
{\bf References}
\begin{description}
\item
Biedermann, S., Dette, H., Zhu, W., 2006.
Optimal designs for dose-response models with restricted design spaces.
J. Amer. Statist. Ass. 101, 747--759.
\item
Ford, I., Torsney, B., Wu, C.F.J., 1992.
The use of a canonical form in the construction of locally optimal designs for nonlinear problems.
J. R. Stat. Soc. Ser B 54, 569--583.
\item
G\"{a}nssler, P., Stute, P., 1977.
Wahrscheinlichkeitstheorie.
Berlin: Springer.
\item
Genz, A., Bretz, F., 2000. Computation of Multivariate Normal and $t$ Probabilities. Berlin: Springer.
\item
Genz, A., Bretz, F., Miwa, T., Mi, X., Leisch, F., Scheipl, F., Hothorn T., 2017. \texttt{mvtnorm}: Multivariate Normal and $t$ Distributions. \texttt{R} package version 1.0-6. URL \texttt{http://CRAN.R-project.org/package=mvtnorm}\,.
\item
Gra{\ss}hoff, U., Gro{\ss}mann, H., Holling, H., Schwabe, R., 2004.
Optimal design for main effects in linear paired comparison models.
J. Statist. Plann. Inference 126, 361-376.
\item
Gra{\ss}hoff, U., Gro{\ss}mann, H., Holling, H., Schwabe, R., 2007.
Design optimality in multifactor generalized linear models in the presence of an unrestricted quantitative factor.
J. Statist. Plann. Inference 137, 3883--3893.
\item
Gra{\ss}hoff, U., Gro{\ss}mann, H., Holling, H., Schwabe, R., 2013.
Optimal design for discrete choice experiments.
J. Statist. Plann. Inference 143, 167--175.
\item
Gro{\ss}mann, H., Schwabe, R., 2015.
Design for discrete choice experiments.
In: Dean, A., Morris, M., Stufken, J., Bingham, D. (Eds.), Handbook of design and analysis of experiments. Boca Raton: CRC Press, a Chapman \& Hall book, pp. 787-832.	
\item
Kanninen, B., 2002. 
Optimal design for multinomial choice experiments.
J. Mark. Res. 39, 214--227.
\item Kiefer, J., 1974.
General Equivalence Theory For Optimum Designs (Approximate Theory).
Ann. Statist. 2, 849--879.
\item
Schwabe, R., 1996.
Optimum Designs for Multi-Factor Models.
New York: Springer.
\item
Silvey, D., 1980.
Optimal Design.
London: Chapman \& Hall.
\item
Sitter, R.R., Torsney, B., 1995.
D-optimal designs for generalized linear models.
In: Kitsos, C.P., M\"uller, W.G. (Eds.),  MODA\,4 - Advances in Model-Oriented Data Analysis.
Heidelberg: Physica, pp 87-102.
\end{description}

\end{document}